\documentclass[aps,prd,twocolumn,superscriptaddress,nofootinbib]{revtex4}

\usepackage{hyperref}
\usepackage{amsmath,amssymb,amsfonts}
\usepackage{color}
\usepackage{graphicx}
\usepackage{enumerate} 
\usepackage{colordvi} 
\usepackage{bm}
\usepackage{multirow}

\newcommand{\be}{\begin{equation}}
\newcommand{\ee}{\end{equation}}
\newcommand{\bey}{\begin{eqnarray}}
\newcommand{\eey}{\end{eqnarray}}
\newcommand{\nn}{\nonumber}
\newcommand{\obs}{{\rm obs}}

\newcommand{\fid}{{\rm fid}}
\newcommand{\bfk}{\boldsymbol{k}}

\newcommand{\bfx}{\boldsymbol{x}}
\newcommand{\bfz}{\boldsymbol{z}}
\newcommand{\bfF}{\boldsymbol{F}}
\newcommand{\bfS}{\boldsymbol{S}}
\newcommand{\bfC}{\boldsymbol{C}}

\newcommand{\DM}{D_{\rm M}}
\newcommand{\DA}{D_{\rm A}}
\newcommand{\Omegam}{\Omega_{\rm m}}
\newcommand{\OmegaK}{\Omega_K}
\newcommand{\OmegaDE}{\Omega_{\rm DE}}
\newcommand{\omegacb}{\omega_{\rm cb}}
\newcommand{\omegab}{\omega_{\rm b}}
\newcommand{\Omegacb}{\Omega_{\rm cb}}
\newcommand{\Omegab}{\Omega_{\rm b}}
\newcommand{\Omegac}{\Omega_{\rm c}}
\newcommand{\AIA}{A_{\rm IA}}

\newcommand{\wt}{\widetilde}
\newcommand{\lin}{{\rm lin}}
\newcommand{\Plin}{P_\lin}
\newcommand{\PEE}{P_{EE}}
\newcommand{\Pgg}{P_{gg}}
\newcommand{\Pvv}{P_{vv}}
\newcommand{\PgE}{P_{gE}}
\newcommand{\PvE}{P_{vE}}
\newcommand{\Pgv}{P_{gv}}
\newcommand{\PmE}{P_{\mu E}}
\newcommand{\wtPgg}{\wt{P}_{gg}}
\newcommand{\wtPEE}{\wt{P}_{EE}}
\newcommand{\wtPvv}{\wt{P}_{vv}}
\newcommand{\Kg}{K_{g}}
\newcommand{\Kv}{K_{v}}
\newcommand{\KE}{K_{E}}
\newcommand{\KB}{K_{B}}
\newcommand{\hmpci}{h\,{\rm Mpc}^{-1}}
\newcommand{\himpc}{h^{-1}\,{\rm Mpc}}
\newcommand{\FoM}{{\rm FoM}}
\newcommand{\Cov}{{\rm Cov}}

\newcommand{\camb}{\texttt{CAMB} }
\newcommand{\slos}{D_{\rm los}}
\newcommand{\Wksz}{G_\parallel}

\begin{document}
\title{
Tightening geometric and dynamical constraints on dark energy and gravity: \\
galaxy clustering, intrinsic alignment and kinetic Sunyaev-Zel'dovich effect
}
\author{Teppei Okumura}\email{tokumura@asiaa.sinica.edu.tw}
\affiliation{Academia Sinica Institute of Astronomy and Astrophysics (ASIAA), No. 1, Section 4, Roosevelt Road, Taipei 10617, Taiwan}
\affiliation{Kavli Institute for the Physics and Mathematics of the Universe (WPI), UTIAS, The University of Tokyo, Kashiwa, Chiba 277-8583, Japan}

\author{Atsushi Taruya}
\affiliation{Yukawa Institute for Theoretical Physics, Kyoto University, Kyoto 606-8502, Japan}
\affiliation{Kavli Institute for the Physics and Mathematics of the Universe (WPI), UTIAS, The University of Tokyo, Kashiwa, Chiba 277-8583, Japan}

\date{\today}

\begin{abstract}
Conventionally, in galaxy surveys, cosmological constraints on the
growth and expansion history of the universe have been obtained from
the measurements of redshift-space distortions and baryon acoustic
oscillations embedded in the large-scale galaxy density field. In this
paper, we study how well one can improve the cosmological constraints
from the combination of the galaxy density field with velocity and
tidal fields, which are observed via the kinetic Sunyaev-Zel'dovich
(kSZ) and galaxy intrinsic alignment (IA) effects, respectively. For
illustration, we consider the deep galaxy survey by Subaru Prime Focus
Spectrograph, whose survey footprint perfectly overlaps with the
imaging survey of the Hyper Suprime-Cam and the CMB-S4 experiment. We
find that adding the kSZ and IA effects significantly improves
cosmological constraints, particularly when we adopt the non-flat cold
dark matter model which allows both time variation of the dark energy
equation-of-state and deviation of the gravity law from general
relativity. Under this model, we achieve $31\%$ improvement for the
growth index $\gamma$ and $>35\%$ improvement for other parameters
except for the curvature parameter, compared to the case of the
conventional galaxy-clustering-only analysis. As another example, we
also consider the wide galaxy survey by the {\it Euclid} satellite, in
which shapes of galaxies are noisier but the survey volume is much
larger. We demonstrate that when the above model is adopted, the
clustering analysis combined with kSZ and IA from the deep survey can
achieve tighter cosmological constraints than the clustering-only
analysis from the wide survey.
\end{abstract}

\maketitle

\flushbottom
\section{Introduction \label{sec:introduction}}

The baryon acoustic oscillations (BAO)
\cite{Peebles:1970,Sunyaev:1970,Eisenstein:1998} and redshift-space
distortions (RSD) \cite{Peebles:1980,Kaiser:1987,Hamilton:1998}
imprinted in large-scale galaxy distribution have been widely used as
powerful tools to constrain the expansion and growth history of the
Universe. Measurements of these signals enable galaxy clustering from
redshift surveys to be one of the most promising probes to clarify the
origin of the late-time cosmic acceleration, which could be explained
by dark energy or modification of gravity
\cite{Peacock:2001,Tegmark:2004,Eisenstein:2005,Okumura:2008,Guzzo:2008,Weinberg:2013,Aubourg:2015,Okumura:2016,Alam:2017,Suyu:2018,Alam:2021}.
Upcoming spectroscopic galaxy surveys, including the Subaru Prime
Focus Spectrograph (PFS) \cite{Takada:2014}, the Dark Energy
Spectroscopic Instrument (DESI) \cite{DESI-Collaboration:2016}, the
{\it Euclid} space telescope
\cite{Laureijs:2011,Capak:2019,Euclid_Collaboration:2020}, and the
Nancy Grace Roman Space Telescope
\cite{Spergel:2013,Spergel:2013a,Eifler:2021}, aim to constrain the
dark energy equation-of-state and deviation of the gravitational law
from general relativity (GR) with a precision at the sub-percent
level.

In order to maximize the information encoded in the galaxy
distribution in the large-scale structure (LSS) and to constrain
cosmological parameters as tightly as possible, one needs to
effectively utilize synergies between galaxy redshift surveys and
other observations.  In this respect, there is a growing interest of
using two effects below as new probes of the LSS to improve
cosmological constraints, complementary to the conventional galaxy
clustering analysis.  The first is the kinetic Sunyaev-Zel'dovich
(kSZ) effect \cite{Sunyaev:1980,Ostriker:1986}, which can be observed
via the measurement of cluster velocities by a synergy between galaxy
surveys and cosmic microwave background (CMB) experiments.
Theoretical and forecast studies suggest that kSZ measurements could
provide robust tests of dark energy and modified gravity theories on
large scales
\cite{Hernandez-Monteagudo:2006,Okumura:2014,Mueller:2015,Sugiyama:2016,Sugiyama:2017,Zheng:2020}.
The kSZ effect has been detected through the cross-correlations of CMB
data with galaxy positions from various redshift surveys
\cite{Hand:2012,De-Bernardis:2017,Sugiyama:2018,Planck-Collaboration:2018,Li:2018,Calafut:2021,Chaves-Montero:2021,Chen:2022}.

The second probe is intrinsic alignment (IA) of galaxy shapes with the
surrounding large-scale matter density field. The IA was originally
proposed as a source of systematic effects on the measurement of the
cosmological gravitational lensing
\cite{Croft:2000,Heavens:2000,Lee:2000,Pen:2000,Catelan:2001,Jing:2002,Hirata:2004,Heymans:2004,Mandelbaum:2006,Hirata:2007,Okumura:2009,Okumura:2009a,Blazek:2011,Singh:2015}.
However, since the spatial correlation of IA follows the gravitational
tidal field induced by the LSS, it contains valuable information and
is considered as a cosmological probe complimentary to the galaxy
clustering
\cite{Schmidt:2012,Faltenbacher:2012,Chisari:2013,Chisari:2016,Okumura:2020,Okumura:2020a,Taruya:2020,Masaki:2020,Akitsu:2021a,Akitsu:2021}.
Ongoing and future galaxy surveys focus on observing LSS at higher
redshifts, $z>1$, at which the emission line galaxies (ELG) would be
an ideal tracer of the LSS
\cite{Laureijs:2011,Takada:2014,Tonegawa:2015,Okada:2016,Dawson:2016,DESI-Collaboration:2016}.
Although IA has not yet been detected for ELG
\cite{Mandelbaum:2006,Mandelbaum:2011,Tonegawa:2018,Tonegawa:2022},
recent work \cite{Shi:2021} has proposed an effective estimator to
determine the IA of dark-matter halos using ELG, enhancing the
signal-to-noise ratio at a statistically significant level.  In any
case, the accurate determination of galaxy shapes is of critical
importance for IA to be a powerful tool to constrain cosmology.  Thus,
the synergy between imaging and spectroscopic surveys is essential
because the accurate galaxy shapes and positions are determined from
the former and latter, respectively.

In this paper, using the Fisher matrix formalism, we simultaneously
analyze the velocity and tidal fields observed by the kSZ and IA
effects, respectively, together with galaxy clustering. The
combination of galaxy clustering with either IA or kSZ has been
studied in earlier studies
\cite[e.g.,][]{Mueller:2015,Sugiyama:2017,Taruya:2020}.  This
is the first joint analysis of these three probes and we want to see
if cosmological constraints can be further improved by combining the
combination.  We emphasize that the question we want to address is not
trivial at all because these probes utilize the information embedded
in the same underlying matter fluctuations.  Nevertheless, a key point
is that these different probes suffer from different systematic
effects, and can be in practice complementary to each other, thus used
as a test for fundamental observational issues, such as the Hubble
tension \cite{Verde:2019}, if the constraining power of each probe is
similar.  Furthermore, analyzing the kSZ and IA simultaneously enables
us to study the correlation of galaxy orientations in phase space as
proposed in our recent series of work
\cite{Okumura:2017a,Okumura:2018,Okumura:2019,Okumura:2020,Okumura:2020a}.
For our forecast, we mainly consider the PFS-like deep galaxy survey
\cite{Takada:2014} which overlaps with the imaging survey of the Hyper
Suprime-Cam (HSC) \cite{Miyazaki:2012,Aihara:2018} and the CMB Stage-4
experiment (CMB-S4) \cite{Abazajian:2016}.  To see how the
cosmological gain by adding the IA and kSZ effects to galaxy
clustering can be different for different survey geometries, we also
analyze the {\it Euclid}-like wide galaxy survey
\cite{Laureijs:2011,Euclid_Collaboration:2020}.

The rest of this paper is organized as follows.  In section
\ref{sec:preliminaries} we briefly summarize geometric and dynamic
quantities to be constrained.  Section \ref{sec:power_spectra}
presents power spectra of galaxy density, velocity and ellipticity
fields and their covariance matrix.  We perform a Fisher matrix
analysis and present forecast constraints in section
\ref{sec:results}, with some details further discussed in section
\ref{sec:discussion}.  Our conclusions are given in section
\ref{sec:conclusion}.  Appendix \ref{sec:prior} describes the CMB
prior used in this paper.  In appendix \ref{sec:conservative}, we
present conservative forecast constraints by restricting the analysis
to large scales where linear perturbation theory is safely applied.


\section{Preliminaries \label{sec:preliminaries}}
\subsection{Distances}

The comoving distance to a galaxy at redshift $z$, $\chi(z)$, is given
by
\be
\chi(z) = \int^z_0 \frac{cdz'}{H(z')} \, , \label{eq:comoving}
\ee
with $c$ being the speed of light. The function $H(z)$ is the Hubble
parameter which describes the expansion rate of the universe. Writing
it as $H(z)=H_0 E(z)$, we define the present-day value of the Hubble
parameter by $H_0\equiv H(z=0)$, which is often characterized by the
dimensionless Hubble constant, $h$, as $H_0=100\,h\,{\rm km~
  s}^{-1}{\rm Mpc}^{-1}$. Then the time-dependent function $E(z)$ is
obtained from the Friedmann equation, and is expressed in terms of the
(dimensionless) density parameters. In this paper, we consider the
universe whose cosmic expansion is close to that in the standard
cosmological model, with the dark energy having the time-varying
equation of state. Allowing also the non-flat geometry, the function
$E(z)$ is given by
\bey
E^2(z) &= & \Omegam (1+z)^{3} + \OmegaK(1+z)^2 \nn \\
&+ & \OmegaDE (1+z)^{3(1+w_0+w_a)} \exp{\left[ -3w_a\frac{z}{1+z} \right]} , \label{eq:friedmann}
\eey
where $\Omegam$, $\OmegaDE$ and $\OmegaK$ are the present-day energy
density fractions of matter, dark energy and curvature, respectively,
with $\Omegam+\OmegaDE+\OmegaK=1$.  In equation (\ref{eq:friedmann}),
the time-varying equation-of-state parameter for dark energy, denoted
by $w(z)$, is assumed to be described by a commonly used and well
tested parameterization \cite{Chevallier:2001,Linder:2003}:
\be
w(z) =w_0+w_a\frac{z}{1+z} = w_0 + w_a(1-a),
\ee
where $a=(1+z)^{-1}$ is the scale factor, and $w_0$ and $w_a$
characterize the constant part and the amplitude of time variation of
the dark energy equation of state, respectively (see e.g.,
Ref. \cite{Colgain:2021}, which studied how the different
parameterization of $w(z)$ affects the constraining power of the
deviation of a cosmological constant.)

The angular diameter distance, $\DA(z)$, is given as
\be
\DA(z)=(1+z)^{-1}\frac{c}{H_0}S_K\left( \frac{\chi(z)}{c/H_0} \right),
\label{eq:angular_diameter_distance}
\ee
where
\begin{align}
S_K(x) = \left\{ 
\begin{array}{lll}
\sin{\left( \sqrt{-\OmegaK}x \right)} / \sqrt{-\OmegaK} &  & \OmegaK<0, \\
x & & \OmegaK=0, \\
\sinh{\left( \sqrt{\OmegaK}x \right)} / \sqrt{\OmegaK}  & & \OmegaK>0. \\
\end{array}
\right.
\end{align}
Negative and positive values of $\OmegaK$ correspond to the closed and
open universe, respectively.  The geometric quantities, $\DA(z)$ and
$H(z)$, are the key quantities we directly constrain from the
measurement of the BAO imprinted in the power spectra.

\subsection{Perturbations}\label{sec:perturbations}

Density perturbations for a given component $i$ ($i=\{\rm m,g\}$ for
matter and galaxies, respectively) are defined by the density contrast
from the mean $\bar{\rho}_i(z)$,
\be
\delta_i(\bfx;z) \equiv \rho_i(\bfx;z) / \bar{\rho}_i(z) - 1. \label{eq:delta}
\ee
Throughout the paper, we assume the linear relation for the galaxy
bias with which the galaxy density fluctuation $\delta_g$ is related
to the matter fluctuation $\delta_m$ through $\delta_g=b_g\,\delta_m$
\cite{Kaiser:1984}. Then, an important quantity to characterize the
evolution of the density perturbation is the growth rate parameter,
defined as
\be
f(z) = -\frac{d\ln D(z)}{d \ln (1+z)} =  \frac{d\ln D(a)}{d \ln a},
\ee
where $D(z)$ is the linear growth factor of the matter perturbation,
$D(z)=\delta_m(\bfx;z) / \delta_m(\bfx;0)$.  The parameter $f$
quantifies the cosmological velocity field and the speed of structure
growth, and thus is useful for testing a possible deviation of the
gravity law from GR \cite{Guzzo:2008}.  For this purpose, it is common
to parameterize the $f$ parameter as
\be
f(z) = \left[ \Omegam(z) \right]^\gamma, \label{eq:gamma}
\ee
where $\Omegam(z)=\Omegam (1+z)^3/ E^{2}(z)$ is the time-dependent
matter density parameter and the index $\gamma$ specifies a model of
gravity, e.g., $\gamma \approx 6/11$ for the case of GR
\cite{Peebles:1980,Linder:2005}.

It is known that a class of modified gravity models exactly follows
the same background evolution as in the $\Lambda$CDM model. However,
the evolution of density perturbations can be different in general
(see, e.g., Ref. \cite{Matsumoto:2020} for degeneracies between the
expansion and growth rates for various gravity models). Thus, it is
crucial to simultaneously constrain the expansion and growth rate of
the universe to distinguish between modified gravity models.

\section{Power spectra and the Fisher matrix}\label{sec:power_spectra}

In this paper, we consider three cosmological probes observed in
redshift space, i.e., density, velocity and ellipticity (tidal)
fields.  While nonlinearity of the density field has been extensively
studied and a precision modeling of its redshift-space power spectrum
has been developed
\cite[e.g.,][]{Peacock:1994,Scoccimarro:2004,Taruya:2010,Okumura:2015},
the understanding of the nonlinearities of velocity and tidal fields
are relatively poor.  However, there are several numerical and
theoretical studies discussed beyond the linear theory, among which a
systematic perturbative treatment has been also exploited (See, e.g.,
Refs. \cite{Okumura:2014,Sugiyama:2016,Sugiyama:2017} and
\cite{Blazek:2015,Blazek:2019,Vlah:2020} for the nonlinear statistics
of velocity and tidal fields, respectively).  It is thus expected that
a reliable theoretical template of their power spectra would be soon
available, and an accessible range of their templates can reach, at
least, at the weakly nonlinear regime. Hence, in our analysis, we
consider the weakly nonlinear scales of $k\leq 0.2\,\hmpci$, as our
default setup.  Nevertheless, in order for a robust and conservative
cosmological analysis, we do not use the shape information of the
underlying matter power spectrum, which contains ample cosmological
information but is more severely affected by the nonlinearities. That
is, our focus in this paper is the measurements of BAO scales and RSD
imprinted in the power spectra, and through the geometric and
dynamical constraints on $D_{\rm A}(z)$, $H(z)$ and $f(z)$, we further
consider cosmological constraints on models beyond the $\Lambda$ cold
dark matter ($\Lambda$CDM) model.  In appendix \ref{sec:conservative},
we perform a more conservative forecast by restricting th analysis to
large scales, $k\leq 0.1\,\hmpci$, where linear perturbation theory
predictions can be safely applied.

In what follows, we discuss how well one can maximize the cosmological
information obtained from the BAO and RSD measurements, based on the
linear theory predictions. While the linear-theory based template is
no longer adequate at weakly nonlinear scales, the signal and
information contained in the power spectrum can be in general
maximized as long as we consider the Gaussian initial condition. In
this respect, the results of our analysis presented below may be
regarded as a theoretical upper bound on the cosmological information
one can get. Furthermore, we assume a plane-parallel approximation for
the cosmological probes \cite{Szalay:1998,Szapudi:2004}, taking the
$z$-axis to be the line-of-sight direction.  While properly taking
into account the wide-angle effect provides additional cosmological
constraints (see, e.g., Refs. \cite{Taruya:2020a},
\cite{Castorina:2020,Shiraishi:2020,Shiraishi:2021} and
\cite{Shiraishi:2021a} for the studies of the wide-angle effects on
density, velocity and ellipticity fields, respectively), we leave the
inclusion of this effect to our analysis as future work.

\subsection{Density, velocity and ellipticity fields}

In this subsection, based on the linear theory description, we write
down the explicit relation between cosmological probes observed in
redshift space to the matter density field. First, the density field
of galaxies in redshift space, which we denote by $\delta_g^S$, is a
direct observable in galaxy redshift surveys, and in Fourier space, it
is related to the underlying density field of matter in real space on
large scales, through $\delta_g^S(\bfk;z) = K_g
(\mu;z)\delta_m(\bfk;z)$.  The factor $\Kg$ is the so-called linear
Kaiser factor given by \cite{Kaiser:1987,Hamilton:1992,Okumura:2011},
\be
\Kg (\mu;z)
=b_g(z)+f(z)\mu^2,
\ee
where $b_g$ is the galaxy bias and $\mu$ is the directional cosine
between the wavevector and line-of-sight direction, $\mu=\hat\bfk
\cdot \hat\bfz$, with a hat denoting a unit vector.  Note that setting
$f$ to zero, the above equation is reduced to the Fourier counterpart
of $\delta_g$ in equation (\ref{eq:delta}).

Next, the cosmic velocity field is related to the density field
through the continuity equation \cite{Fisher:1995,Strauss:1995}.  The
observable through the kSZ effect is the line-of-sight component of
the velocity, $v_\parallel$, and in linear theory, we have (in Fourier
space) $v_\parallel (\bfk;z)= if(z)\mu aH\delta_m(\bfk;z)/k$. To be
precise, the kSZ effect measures the temperature distortion of CMB,
$\delta T$, detected at the position of foreground galaxies.  It is
explicitly written in Fourier space as $\delta T (\bfk;z)=(T_0\tau/c)
v_\parallel (\bfk;z) = i \Kv (\bfk;z) \delta_m(k;z)$, where
\be
\Kv(k,\mu;z) =  \frac{T_0\tau }{c}\frac{f(z)\mu aH(z)}{k} ~ , \label{eq:kv}
\ee
with $\tau$ being the optical depth.  Since the distance to tracers of
the velocity field is measured by redshift, the observed velocity
field is affected by RSD, similarly to the density field in redshift
space.  Unlike the density field, however, the RSD contribution to the
redshift-space velocity field appears at higher order
\cite{Okumura:2014}.  Thus, at leading order, the velocity field
traced in redshift space coincides with that in real space in
linearized theory, $v_\parallel ^S = v_\parallel$.  Note that the kSZ
effect, which appears as secondary CMB anisotropies, is given by a
line-of-sight integral of the velocity field, and thus the expression
of Eq. (\ref{eq:kv}) is just an approximation. We discuss the validity
of this approximation in section \ref{sec:los_ksz}.

An alternative way to measure the velocity field $v_\parallel$ without
observing the temperature distortion is to use velocity surveys, which
enable us to uniquely constrain $f(z)$ \cite{Strauss:1995}.  We,
however, do not consider observables from peculiar velocity surveys.
The main reason is that these observations are limited to the nearby
universe ($z\approx 0$) while we consider joint constraints with other
probes from a single observation of the LSS. Thus, throughout this
paper we refer the velocity field as the temperature distortion
$\delta T$.

\begin{table*}[bt!]
\caption{Statistics and their abbreviations considered for given
  probes.  Note that when two fields, $A$ and $B$, are considered, we
  use not only the auto-correlations ($P_{AA}$ and $P_{BB}$) but also
  the cross correlation, $P_{AB}$.  }
\begin{center}
\begin{tabular}{ l  l c c c c}
\hline 
\hline 
\multirow{2}{*}{Probes} &  \multirow{2}{*}{Statistics} & \multirow{2}{*}{Abbreviations}
 & No. of & \multicolumn{2}{c}{Parameters $\{\theta_\alpha\}$} \\
\cline{5-6}
    &    &    & parameters $N_\theta$ & nuisance    & geometric/dynamical \\
\hline
Clustering    &  $\Pgg$   & $g$ & 4 & $b\sigma_8$&$f\sigma_8,H,\DA$  \\
kSZ  &  $\Pvv$   & $v$ & 4& $\tau$&$f\sigma_8,H,\DA$ \\
IA   &  $\PEE$   & $E$ & 3 &$\AIA$&$H,\DA$ \\
Clustering+IA  &  $\Pgg+\PEE+\PgE$   & $g+E$ &  5 & $b\sigma_8,\AIA$&$f\sigma_8,H,\DA$ \\
Clustering+kSZ  &  $\Pgg+\Pvv+\Pgv$  & $g+v $ & 5 & $b\sigma_8,\tau$&$f\sigma_8,H,\DA$ \\
IA+kSZ  &  $\PEE+\Pvv+\PvE$   & $v+E$ & 5 & $\AIA,\tau$&$f\sigma_8,H,\DA$ \\
Clustering+IA+kSZ  &  $\Pgg+ \PEE+\Pvv +\PgE+\Pgv  +\PvE$
& $g + v + E$ & 6 & $b\sigma_8,\AIA,\tau$&$f\sigma_8,H,\DA$ \\
\hline
\end{tabular}
\label{tab:statistics}
\end{center}
\end{table*}

Finally, we use ellipticities of galaxies as a tracer of the tidal
field.  The two-component ellipticity of galaxies is defined as
\be
\gamma_{(+,\times)}(\bfx;z) = \frac{1-q}{1+q} \left( \cos{(2\phi_x)},\sin{(2\phi_x)} \right),
\ee
where $\phi_x$ is the position angle of the major axis relative to the
reference axis, defined on the plane normal to the line-of-sight
direction, and $q$ is the minor-to-major axis ratio of a galaxy
shape. We set $q$ to zero for simplicity \cite{Okumura:2009}.  As a
tracer of LSS, a leading-order description of the ellipticity field is
to relate $\gamma_{(+,\times)}$ linearly to the tidal gravitational
field, known as the linear alignment (LA) model
\cite{Catelan:2001,Hirata:2004,Okumura:2019,Okumura:2020}. In Fourier
space, this is given by
\be
\gamma_{(+,\times)}(\bfk;z) = b_K(z) \left( k_x^2-k_y^2, 2k_xk_y \right)\frac{\delta_m(\bfk;z)}{k^2}. 
\label{eq:gamma_+,x}
\ee
Just like the velocity field, the ellipticity field is not affected by
RSD in linear theory \cite{Okumura:2020}.  We then define E-/B-modes,
$\gamma_{(E,B)}$, which are the rotation-invariant decomposition of the
ellipticity field \cite{Crittenden:2002},
\bey
\gamma_E(\bfk;z) + i\gamma_B(\bfk;z) = e^{-2i\phi_k}\left\{  \gamma_{+}(\bfk;z)+i\gamma_{\times}(\bfk;z) \right\},
\label{eq:gamma_E_deltam}
\eey
where $\phi_k$ is the azimuthal angle of the wavevector projected on
the celestial sphere (Note that $\phi_k$ has nothing to do with the
directional cosine of the wavevector, and thus $\phi_k \neq
\cos^{-1}{\mu}$).  By writing $\gamma_{(E,B)}(\bfk ; z) =
K_{(E,B)}(\mu;z)\delta_m(\bfk;z)$, we have $\KB=0$ and
\be
\KE(\mu;z) = b_K(z) (1-\mu^2).
\ee
In Eq.~(\ref{eq:gamma_+,x}) or (\ref{eq:gamma_E_deltam}), the
parameter $b_K$ quantifies the response of individual galaxy shapes to
the tidal field of LSS, and it is conventionally characterized by
introducing the parameter $\AIA$ as follows
\cite[e.g.,][]{Schmitz:2018,Kurita:2021}:
\be
b_K(z) = 0.01344 \AIA(z)\Omegam/D(z).
\ee
Note that the parameter $\AIA$ generally depends on properties of the
given galaxy population as well as redshift.  The analysis of
numerical simulations, however, demonstrated that for fixed
galaxy/halo properties, $\AIA$ is nearly redshift-independent
\cite{Kurita:2021}.  We thus treat $\AIA$ as a constant throughout
this paper.

\subsection{Linear power spectra of the three fields} 

As summarized in the previous subsection, the three cosmological
fields, i.e. density, velocity and ellipticity, are related to the
matter field linearly through the coefficients, $\Kg$, $\Kv$ and
$\KE$, respectively.  Provided their explicit expressions, we can
analytically compute the auto-power spectra of these fields and their
cross-power spectra. There are in total six power spectra measured in
redshift space, each of which exhibits anisotropies characterized by
the $\mu$ dependence
\cite{Ballinger:1996,Seo:2003,Matsubara:2004,Okumura:2020}.  Writing
these spectra as $P_{ij}(\bfk;z)=P_{ij}(k,\mu;z)$ with $i,j =
\{g,v,E\}$, they are expressed in a concise form as,
\begin{align}
P_{ij}(k,\mu;z) & =  K_i(k,\mu;z) K_j(k,\mu;z) \Plin(k;z) , \label{eq:Pij}
\end{align}
where $\Plin(k;z) $ is the linear power spectrum of matter fluctuation
in real space. The normalization of the density fluctuation is
characterized by the $\sigma_8$ parameter, defined by the linear RMS
density fluctuation within a sphere of radius $8h^{-1}$Mpc, and thus
$\Plin(k;z) \propto \sigma_8^2(z)$.  While each of the three
auto-power spectra, $\Pgg$, $\Pvv$ and $\PEE$, can be measured from
each of the three individual probes, namely galaxy clustering, kSZ and
IA, respectively, the cross-power spectra become measurable only when
two probes are simultaneously made available.\footnote{Note that this
  terminology is different from that used in past studies: while in
  this paper the kSZ and IA power spectra stand for only $\Pvv$ and
  $\PEE$, respectively, the past studies included the cross-power
  spectrum with density field, $\Pgv$ and $\PgE$, into kSZ and IA
  spectra.}
Particularly, the correlation between velocity and ellipticity fields,
$\PvE$, has been proposed recently by our earlier studies and it can
be probed by the joint analysis of the kSZ (or peculiar velocities)
and IA effects
\cite{Okumura:2017a,Okumura:2018,Okumura:2019,Okumura:2020,Okumura:2020a,van_Gemeren:2020}.
Table \ref{tab:statistics} summarizes all the statistics used in this
paper.

To measure the power spectra, the observed galaxy positions measured
with redshift and angular position need to be converted into the
comoving positions by introducing a reference cosmology, with a help
of equations (\ref{eq:comoving}) and
(\ref{eq:angular_diameter_distance}). An apparent mismatch between the
reference and true cosmology causes a geometric distortion in the
measured power spectra, which is yet another anisotropy known as the
Alcock-Paczynski (AP) effect \cite{Alcock:1979}.  This AP effect has
been extensively investigated for the galaxy power spectrum in
redshift space
\cite{Ballinger:1996,Matsubara:1996,Seo:2003,Blazek:2014}.  The AP
effect on the kSZ and IA statistics has been studied relatively
recently by Refs. \cite{Sugiyama:2017} and \cite{Taruya:2020},
respectively.  In all of the six power spectra, $P_{ij}$, their
observable counterpart $P_{ij}^{\rm obs}$ are related to the true ones
through the relation,
\bey
P_{ij}^{\obs}\left(k_\perp^\fid,k_\parallel^{\fid};z \right) = 
\frac{H(z)}{H^\fid(z)}\left\{ \frac{D_{\rm A}^\fid(z)}{\DA(z)} \right\}^2
P_{ij}\left(k_\perp,k_\parallel ; z \right), 
\label{eq:Pij_AP}
\eey
where $k_\perp$ and $k_\parallel$ are the wavenumber perpendicular and
parallel to the line of sight, $(k_\perp,k_\parallel) =
k(\sqrt{1-\mu^2},\mu)$. The quantities $D_{\rm A}^\fid(z)$ and
$H^\fid(z)$ are the angular diameter distance and expansion rate
computed from fiducial cosmological parameters in the reference
cosmology, and $k_\parallel^\fid=k_\parallel H^\fid(z)/H (z)$ and
$k_\perp^\fid=k_\perp\,\DA(z) / D_{\rm A}^\fid(z)$.  The prefactor
$\frac{H(z)}{H^\fid(z)}\left\{ \frac{D_{\rm A}^\fid(z)}{\DA(z)}
\right\}^2$ accounts for the difference in the cosmic volume in
different cosmologies.

As formulated above, $\Kg$, $\Kv$ and $\KE$ respectively contain two
$(b, f)$, two $(\tau, f)$, and one $(\AIA)$ parameters, and all the
power spectra depend on $(H,\DA)$ through the AP effect (see table
\ref{tab:statistics}).  Thus, we have six parameters in total,
$\theta_\alpha = (b\sigma_8, \AIA\sigma_8,\tau, f\sigma_8,H,\DA)$,
among which the first three are nuisance parameters that we want to
marginalize over.  The latter three parameters carry the cosmological
information which characterize the growth of structure and geometric
distances, and are determined by measuring the anisotropies in the
power spectra.

\subsection{Covariance matrix}\label{sec:covariance}

\begin{figure*}[t]
\centering
\includegraphics[width=0.999\textwidth]{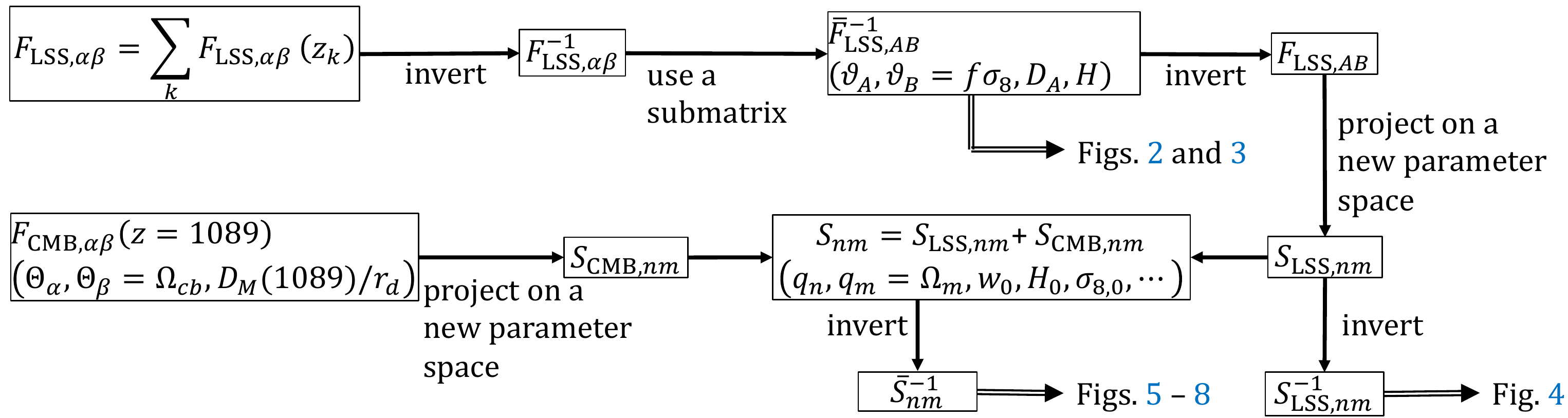}
\caption{Flowchart of our Fisher matrix analysis from dynamical and
  geometric constraints to cosmological parameter constraints. The
  Fisher matrices of the LSS probes, namely galaxy clustering, IA and
  kSZ, are given for each redshift bin $z_k$ at the upper left. The
  Fisher matrix from the CMB prior is given at the lower left. }
\label{fig:flowchart}
\end{figure*}

Writing all the power spectra obtained from the galaxy clustering, kSZ
and IA as \\ $P_{a} = (\Pgg, \PEE, \Pvv, \PgE, \Pgv, \PvE)$, we will
below examine several forecast analysis with a different number of
power spectra, which we denote by $N_P$. Specifically, depending on
how many probes are simultaneously available, we consider seven
possible cases with $N_P=1$, $3$ or $6$, summarized in table
\ref{tab:statistics}. Correspondingly, the covariance matrix
$\Cov_{ab}$ becomes a $N_P \times N_P$ matrix, defined as
$\Cov_{ab}(k,\mu;z) = \left\langle P_aP_b\right\rangle-\left\langle
P_a\right\rangle\left\langle P_b\right\rangle$, for a given
wavevector, ${\bf k}=(k,\mu)$.  The full $6\times 6$ Gaussian
covariance matrix reads
\begin{widetext}
\begin{align}
\Cov_{ab}(k,\mu;z) =
\left[
\begin{array}{cccccc}
2 \{ \wtPgg\}^2  & 2 \{ \PgE\}^2     & 2 \{ \Pgv\}^2   & 2\wtPgg\PgE                         & 2\wtPgg\Pgv                        &2\Pgv\PgE                           \\ 
2 \{ \PgE\}^2    &2 \{ \wtPEE\}^2  &2\{\PvE\}^2     & 2\PgE\wtPEE                         & 2\PgE\PvE                           &2\wtPEE\PvE                       \\ 
2 \{ \Pgv\}^2     &2\{\PvE\}^2       &2 \{ \wtPvv\}^2 & 2\Pgv\PvE                             & 2\wtPvv\Pgv                         & 2\PvE\wt{P}_{vv}                 \\ 
2\wtPgg\PgE    &2\PgE\wtPEE   &2\Pgv\PvE       & \wtPgg \wtPEE + \{ \PgE\}^2  & \wtPgg\PvE + \PgE\Pgv       & \Pgv\wtPEE + \PgE\PvE      \\ 
2\wtPgg\Pgv     &2\PgE\PvE       &2\wtPvv\Pgv   & \wtPgg\PvE + \PgE\Pgv          & \wtPgg \wtPvv + \{\Pgv\}^2 & \PgE\wtPvv + \Pgv\PvE        \\ 
2\Pgv\PgE        &2\wtPEE\PvE   &2\PvE\wtPvv   & \Pgv\wtPEE + \PgE\PvE         & \PgE\wtPvv + \Pgv\PvE      & \wtPEE \wtPvv + \{ \PvE\}^2 \\ 
\end{array}
\right] ~,
\label{eq:covariance}
\end{align}
\end{widetext}
where $\wt{P}_{ii}=\wt{P}_{ii}(k,\mu;z)$ denotes an auto-power
spectrum ($i=\{g,v,E\}$) including the shot noise. Assuming the
Poisson shot noise, we have
\begin{align}
&\wtPgg = \Pgg + \frac{1}{n_g} , 
\\
&\wtPvv = \Pvv + \left( 1 + R_N^2 \right)\left(\frac{T_0 \tau}{c}\right) ^2\frac{(faH\sigma_d)^2}{n_v}, 
\label{eq:Pvv_with_shot_noise}
\\
&\wtPEE = \PEE + \frac{\sigma_\gamma^2}{n_\gamma}, 
\label{eq:PEE_with_shot_noise}
\end{align}
where the quantities $n_g$, $n_v$ and $n_\gamma$ are the number
density of the galaxies obtained from galaxy clustering, kSZ and IA
observations, respectively.  Though different notations are explicitly
used for these three samples, $n_g=n_v=n_\gamma$ when one considers a
single galaxy population for the analysis.  When one uses a single
galaxy population as a tracer of the density, velocity and ellipticity
fields, there should be a shot noise contribution in the cross
correlations. Such a noise term, however, vanishes because $\langle
v_\parallel \rangle =0$ \cite{Sugiyama:2017} and $\langle
\gamma_E\rangle =0$ \cite{Taruya:2020}.

In the shot noise terms of $\Pvv$ and $\PEE$, there appear factors
$\sigma_d = \sqrt{\langle v_\parallel ^2 \rangle}$ and $\sigma_\gamma
= \sqrt{\langle \gamma_E^2 \rangle}$, which respectively represent the
velocity dispersion and shape noise of galaxies, respectively.  Using
perturbation theory, $\sigma_d$ can be evaluated as
\be
\sigma_d^2 = \frac{1}{3}\int\frac{d^3q}{(2\pi)^3}\frac{P_{\theta\theta}(q;z)}{q^2}
= \frac{1}{6\pi^2}\int dq P_{\theta\theta}(q;z)
, \label{eq:sigma_d}
\ee
where $P_{\theta\theta}$ is the power spectrum of velocity
divergence. In the limit of linear theory, we have
$P_{\theta\theta}=P_{\lin}$, and in the standard cosmological model,
it is predicted to give $aH\sigma_{d,\lin} \simeq 600D(z)~{\rm
  km}/{\rm s}$ \cite{Vlah:2012}, and hence $faH\sigma_{d,\lin} \simeq
600f(z)D(z)\approx 300~{\rm km}/{\rm s}$ over the redshift considered
in this work.  Finally, the parameter $R_N$ is the inverse
signal-to-noise ratio of the kSZ temperature fluctuations
\cite{Sugiyama:2017}.  The rms noise for the kSZ measurement of the
CMB-S4 experiment is $\left\langle \delta T\right\rangle \sim 2 \mu
K$, leading to $R_N\sim 10$ \cite{Sugiyama:2017}.

Note that considering only the Gaussian contribution of the covariance
matrix (equation (\ref{eq:covariance})) may underestimate the
statistical errors. Particularly, the kSZ effect generally suffers
from a correlated non-Gaussian noise due to the residual foreground
contamination, e.g., cosmic infrared background and thermal SZ effect
\cite[see e.g.,][]{Bobin:2014,Planck-Collaboration:2018}. Though our
focus is on relatively large scales and we adopt the Gaussian
covariance, such non-Gaussian contributions need to be taken into
account for a more realistic forecast study.

\subsection{Fisher matrix formalism}\label{sec:fisher}

To quantify the constraining power for the dynamical and geometric
parameters above and cosmological parameters, we use the Fisher matrix
formalism.  Although forecast studies with the Fisher matrix have been
widely performed in cosmology, there is a limited number of relevant
works that consider the kSZ and IA observations to constrain
cosmology, specifically through the RSD and AP effect. One is the
paper by Sugiyama, Okumura \& Spergel \cite{Sugiyama:2017}, who
discussed a benefit of using kSZ observations. Another paper is Taruya
\& Okumura \cite{Taruya:2020}, who demonstrated that combining galaxy
clustering with IA observations is beneficial and improves geometric
and dynamical constraints. The present paper complements these two
previous works, and further put forward the forecast study by
combining all three probes.

Given a set of parameters to be estimated, $\{\theta_\alpha\}$, where
$\alpha=1,\cdots,N_\theta$, and provided a set of observed power
spectra $\{P_a\}$, the Fisher matrix is evaluated with
\begin{align}
F_{\alpha\beta} = & \frac{V_s}{4\pi^2} \int^{k_{\rm max}}_{k_{\rm min}} dk k^2 \int ^{1}_{-1}d\mu \nn \\
& \times \sum_{a,b=1}^{N_P} \frac{\partial P_a(k,\mu)}{\partial\theta_\alpha}
\left[ \Cov^{-1}\right]_{ab}\frac{\partial P_b(k,\mu)}{\partial\theta_\beta},
\label{eq:Fisher_matrix}
\end{align}
where $V_s$ is the comoving survey volume for a given redshift range,
$z_{\min} \leq z \leq z_{\max}$, and $k_{\rm min}$ and $k_{\rm max}$
are respectively the minimum and maximum wavenumbers used for
cosmological data analysis, the former of which is specified with the
survey volume by $k_{\rm min}=2\pi/V_s^{1/3}$.  Note that for the
analysis using a single probe ($N_P=1$), namely when we consider
either of $\Pgg$, $\PEE$ or $\Pvv$, the covariance matrix ${\rm
  Cov}_{ab}$ is reduced to the power spectrum squared (see equation
(\ref{eq:covariance})).

Provided the Fisher matrix, the expected errors on the parameters of
interest, marginalizing over other parameters, are computed by
inverting the Fisher matrix and constructing the
$\overline{N}_{\vartheta}\times\overline{N}_{\vartheta}$ submatrix
$\overline{\bfF}$; for example, when one wants to evaluate the
two-dimensional error contours for a specific pair of parameters,
$\vartheta_A \in\,\theta_\alpha$ $(A=1,2)$, the $2\times2$ submatrix
is constructed with $C_{AB} \equiv [ \overline{\bfF} ]^{-1}_{AB} $
$(A,B=1,2)$. Also, the one-dimensional marginalized error on a
parameter $\vartheta_A$ is obtained from $\sigma_A^2 \equiv [
  \overline{\bfF} ]^{-1}_{AA}$ (see, e.g., Ref. \cite{Seo:2003} for
details).

Although our original Fisher matrix is given for the parameters
$\{\theta_\alpha\}$ determined from the AP effect and RSD, the
model-independent geometric and dynamical constraints are translated
into specific cosmological model constraints by projecting the matrix
into a new parameter space of interest,
\be
S_{nm} = \sum_{\alpha,\beta}^{N_\theta}\frac{\partial\theta_\alpha}{\partial q_n} F_{\alpha\beta} \frac{\partial\theta_\beta}{\partial q_m},
\label{eq:projection_fisher}
\ee
where $\{q_n\}$ is the set of parameters in the new parameter space
($n=1,\cdots,N_q$), i.e., non-flat $w_0w_a\gamma$ CDM model and others
in our case (see section \ref{sec:constraints1}), and $\bfS$ is thus a
$N_q\times N_q$ matrix.  Once again, the uncertainties of the
parameters can be obtained by taking the submatrix, e.g., $C_{AB}
\equiv [ \overline{\bfS} ]^{-1}_{AB} $, $\sigma_{A}^2 \equiv [
  \overline{\bfS} ]^{-1}_{AA}$, etc.

For a further discussion on the performance of the constraining power
on multiple parameters, we compute the Figure-of-merit (FoM) defined
by
\be
{\rm FoM} = \left\{{\rm det}(\overline{\bfF})\right\}^{1/\overline{N}_\vartheta}, \ \ \ \ \ \ \ 
{\rm FoM} = \left\{{\rm det}(\overline{\bfS})\right\}^{1/\overline{N}_\varrho} \label{eq:FoM},
\ee
where quantities with the bar, $\overline{\bfF}$ and
$\overline{\bfS}$, denote $\overline{N}_\vartheta \times
\overline{N}_\vartheta$ and $\overline{N}_\varrho \times
\overline{N}_\varrho$ submatrices of $\bfF$ and $\bfS$
($\overline{N}_\vartheta < N_\theta, \overline{N}_\varrho<N_q$),
respectively, constructed through the inversion described above.  In
the definition provided in Ref. \cite{Albrecht:2006},
$\overline{N}_\vartheta = \overline{N}_\varrho = 2$ and the obtained
FoM describes the inverse of the area of the error contour in the
marginalized parameter plane for two parameters.  Here, the FoM is
defined for an arbitrary number of parameters, and the obtained value
corresponds to a mean radius of the $\overline{N}_\vartheta$ (or
$\overline{N}_\varrho$) dimensional volume of the errors.


\begin{table}[bt!]
\caption{Expected volume, number density and bias of emission line
  galaxies for given redshift ranges, $z_{\min}\leq z \leq z_{\max}$
  of the deep (PFS-like) survey, taken from Ref. \cite{Takada:2014}.}
\begin{center}
\begin{tabular}{ cc  ccc  }
\hline \hline 
\multicolumn{2}{ c }{Redshift} &  Volume $V_s$                        &  $10^4n $ &  Bias \\
$z_{\rm min} $ &  $z_{\rm max}$   &  ($h^{-3}{\rm Gpc}^3$) &  $(h^3{\rm Mpc}^{-3})$ & $b_g$  \\
\hline
0.6&0.8 & $0.59$ & $1.9$ & $1.18$  \\  
0.8&1.0 & $0.79$ & $6.0$ & $1.26$  \\  
1.0&1.2 & $0.96$ & $5.8$ & $1.34$  \\  
1.2&1.4 & $1.09$ & $7.8$ & $1.42$  \\  
1.4&1.6 & $1.19$ & $5.5$ & $1.50$  \\  
1.6&2.0 & $2.58$ & $3.1$ & $1.62$  \\  
2.0&2.4 & $2.71$ & $2.7$ & $1.78$  \\  
\hline \hline
\end{tabular}
\label{tab:survey_pfs}
\end{center}
\end{table}

\section{Results}\label{sec:results}

In this section, we present geometric and dynamical constraints on
cosmological parameters based on the Fisher matrix analysis of galaxy
clustering, IA and kSZ effects.  In figure \ref{fig:flowchart}, we
summarizes the steps of the analysis of this section graphically,
motivated by figure 2 of Ref. \cite{Seo:2003}.

\subsection{Setup}\label{sec:setup}

To jointly analyze the galaxy clustering, IA and kSZ, we need to use
data from galaxy surveys and CMB experiments: positions and shapes of
galaxies are respectively used to quantify clustering and IA from a
galaxy survey, while the velocity field is inferred by observing the
CMB temperature distortion at the angular position of each galaxy.


\begin{table}[bt!]
\caption{ Same as table \ref{tab:survey_pfs} but for the wide ({\it
    Euclid}-like) survey, taken from
  Ref. \cite{Euclid_Collaboration:2020}.}
\begin{center}
\begin{tabular}{ cc  ccc  }
\hline \hline 
\multicolumn{2}{ c }{Redshift} &  Volume $V_s$                    &  $10^{4}n $ & Bias  \\
$z_{\rm min} $ &  $z_{\rm max}$  &  ($h^{-3}{\rm Gpc}^3$) &  $(h^3{\rm Mpc}^{-3})$ & $b_g$  \\
\hline
0.9&1.1 & $7.94$ & $6.86$ & $1.46$  \\  
1.1&1.3 & $9.15$ & $5.58$ & $1.61$  \\  
1.3&1.5 & $10.05$ & $4.21$ & $1.75$  \\  
1.5&1.8 & $16.22$ & $2.61$ & $1.90$  \\ 
\hline \hline
\end{tabular}
\label{tab:survey_euclid}
 \end{center}
\end{table}

\begin{figure*}[t]
\centering
\includegraphics[width=0.43\textwidth]{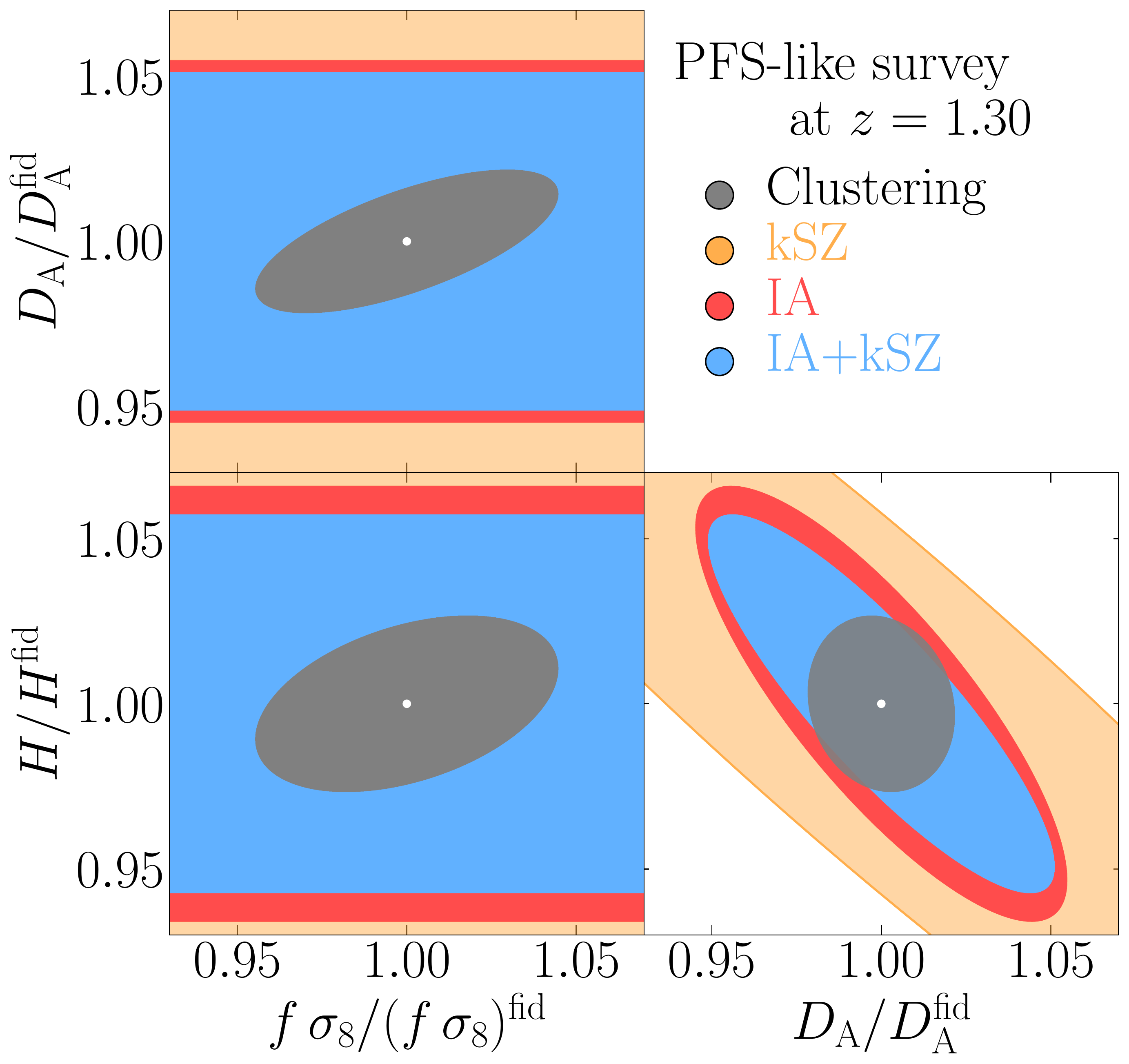}
\ \ \ \ \ \ 
\includegraphics[width=0.43\textwidth]{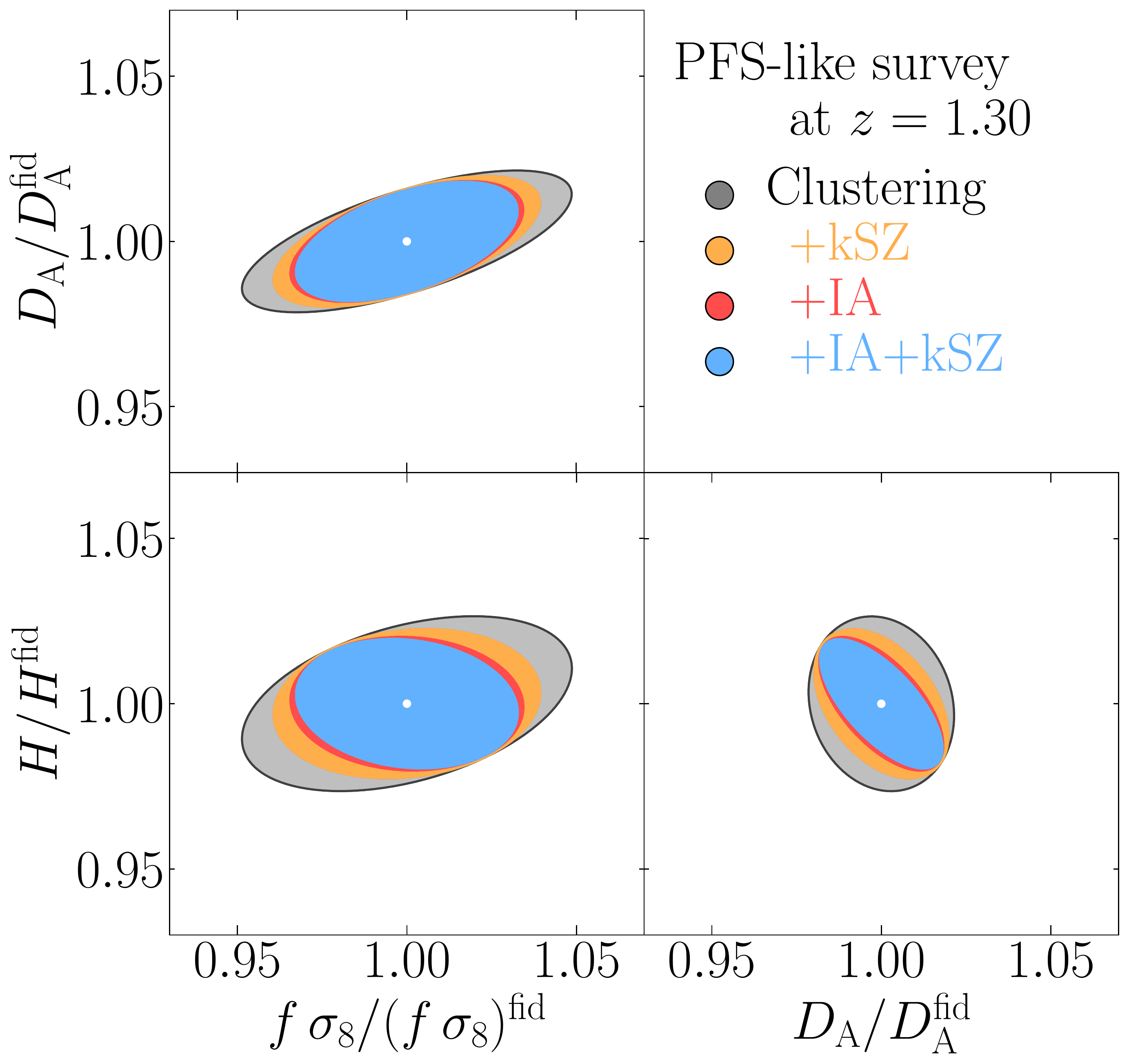}
\caption{ Two-dimensional $1\sigma$ error contours on the geometric
  distances, $\DA(z)$ and $H(z)$, and the linear growth rate
  $f(z)\sigma_8(z)$, expected from the wide (PFS-like) survey.  Since
  there are seven redshift bins, we here show the result for the
  central redshift bin, $1.2<z<1.4$, as an example.  {\it Left panel}:
  Constraints from each of the clustering, kSZ and IA, and the
  combination of the latter two.  {\it Right panel}: Similar to the
  left panel but joint constraints from the combination of kSZ, IA and
  galaxy clustering. Note that the joint constraints obtained from
  clustering and IA (red contours) almost overlap with those from
  clustering, IA and kSZ (blue contours).  }
\label{fig:2D_errors_PFS_f_dA_H}
\end{figure*}

As we mentioned in section \ref{sec:introduction}, there are a number
of planned spectroscopic galaxy surveys aiming at constraining
cosmology with a high precision. These surveys are generally
categorized into the two types: (narrow but) deep surveys and (shallow
but) wide surveys. In the Fisher matrix analysis below, we consider
the Subaru PFS and {\it Euclid} as examples of deep and wide surveys,
respectively, both of which target emission line galaxies (ELG) as a
tracer of the LSS. Tables \ref{tab:survey_pfs} and
\ref{tab:survey_euclid} show the redshift range, survey volume, and
number density and bias of the ELG samples for the PFS
\cite{Takada:2014} and {\it Euclid} \cite{Euclid_Collaboration:2020},
respectively.  Ref. \cite{Shi:2021} has proposed an estimator to
directly detect IA of host halos using the observation of the ELGs. In
the forecast analysis presented below, we consider that the power
spectra related to the IA are measured with this
estimator.\footnote{Even though we use elliptical galaxies as a tracer
  of the tidal field as in the conventional analysis, we can present a
  similar analysis based on luminous red galaxy samples from, i.e.,
  DESI, and the main results below will not change qualitatively
  (e.g., \cite{Taruya:2020}).}
Following the result of Ref. \cite{Shi:2021}, we set the fiducial
value of the IA amplitude to $\AIA=18$, assuming its redshift
independence.  The PFS galaxy sample provides high-quality shape
information thanks to the imaging survey of the HSC
\cite{Miyazaki:2012,Aihara:2018}, and we thus set the shape noise,
$\sigma_\gamma$, to $\sigma_\gamma=0.2$ for the deep survey
\cite{Hikage:2019}.  For the wide survey, following
Ref.~\cite{Euclid_Collaboration:2020}, we set it to
$\sigma_\gamma=0.3$.  We will discuss the effect of changing the
fiducial values of $\AIA$ and $\sigma_\gamma$ in section
\ref{sec:discussion}.

Similarly to the forecast study of the kSZ effect in
Ref. \cite{Sugiyama:2017}, we consider CMB-S4 \cite{Abazajian:2016} as
a CMB experiment for the expected observation of the kSZ effect.
While the angular area of the PFS is completely overlapped with that
of the CMB-S4, the half of the {\it Euclid} area is covered by the
CMB-S4 \cite{Euclid_Collaboration:2022}.  Thus, when considering the
statistics related to the kSZ effect, namely $\Pvv$, $\Pgv$ and
$\PvE$, in the wide survey, the elements of the covariance matrix for
these statistics are multiplied by two.  Furthermore, the values of
$k_{\min}$ for these terms become larger by the factor of $2^{1/3}$.
We choose $R_{\rm N}=10$ as our fiducial choice, following
Ref. \cite{Sugiyama:2017}.  For the velocity dispersion, we use the
liner theory value as a fiducial value, $\sigma_v = \sigma_{v,{\rm
    lin}}$. The combination of $(1+R_{\rm N}^2)\sigma_v^2$ contributes
to the shot noise of the kSZ power spectrum.  We will test the effect
of these choices in section \ref{sec:discussion}.

In the following analysis, we assume the spatially flat $\Lambda$CDM
model as our fiducial model \cite{Planck-Collaboration:2016}: $\Omegam
= 1-\OmegaDE = 0.315$, $\OmegaK=0$, $w_0 = -1$, $w_a=0$,
$H_0=67.3~[{\rm km/s/Mpc}]$ and the present-day value of $\sigma_8$,
$\sigma_{8,0}\equiv\sigma_8(z=0)$, to be $\sigma_{8,0} = 0.8309$.  For
computation of the linear power spectrum in equation (\ref{eq:Pij}),
$\Plin(k;z)$, we use the publicly-available \camb code
\cite{Lewis:2000}.  When we consider the model which allows deviation
of the structure growth from GR prediction, we set the fiducial value
of $\gamma$ in equation (\ref{eq:gamma}) to be consistent with GR,
$\gamma=0.545$.

Finally, the maximum wavenumber of the power spectra used for the
cosmological analysis with the Fisher matrix is set to $k_{\rm
  max}=0.2\,h$\,Mpc$^{-1}$. While forecast results with this choice,
presented below as our main results, give tight geometrical and
dynamical constraints, we also consider in Appendix
\ref{sec:conservative} a conservative choice of $k_{\rm
  max}=0.1\,h\,{\rm Mpc}^{-1}$, and discuss its impact on the
parameter constraints.


\begin{figure*}[bt]
\centering
\includegraphics[width=0.49\textwidth]{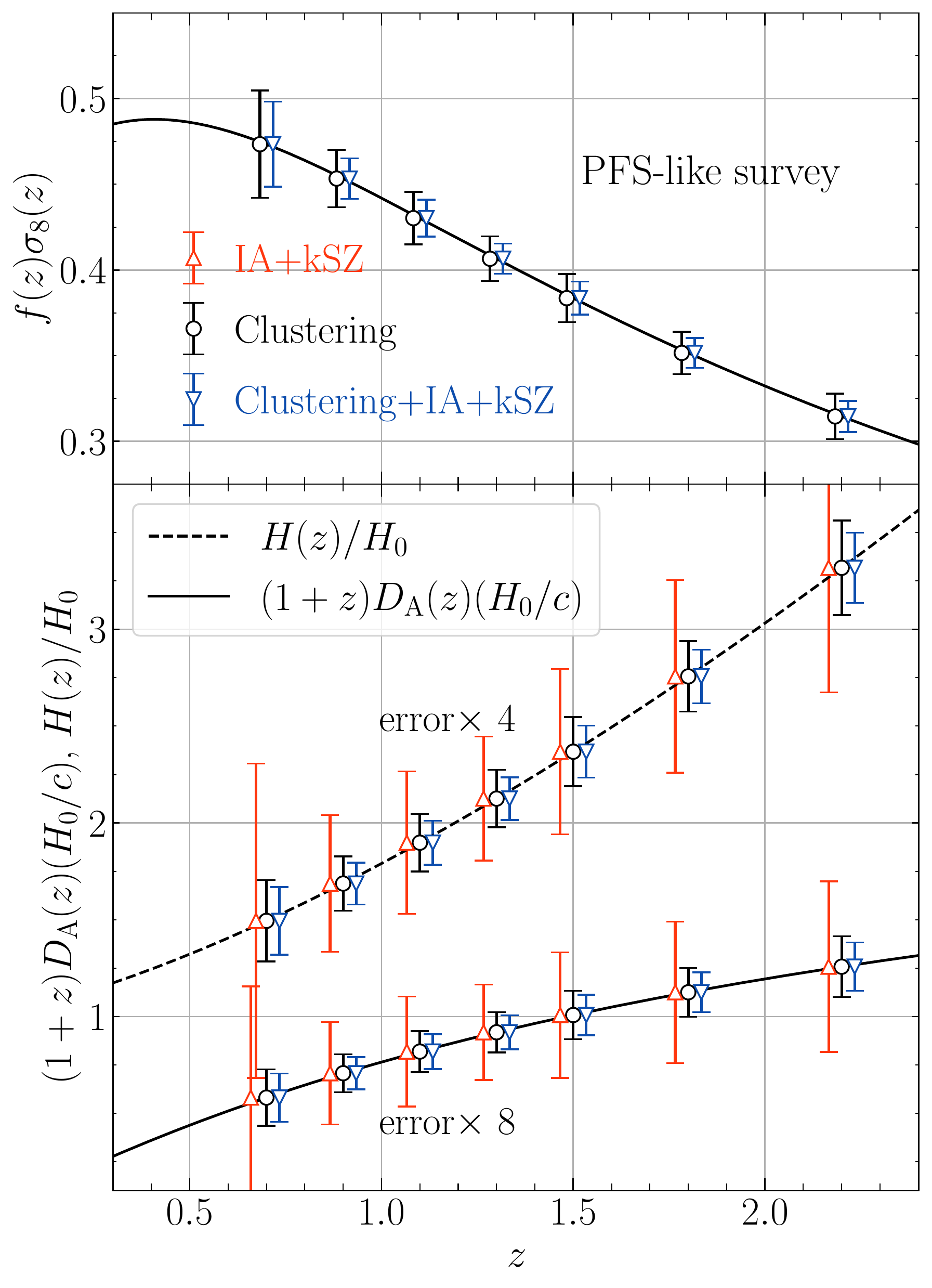}
\includegraphics[width=0.49\textwidth]{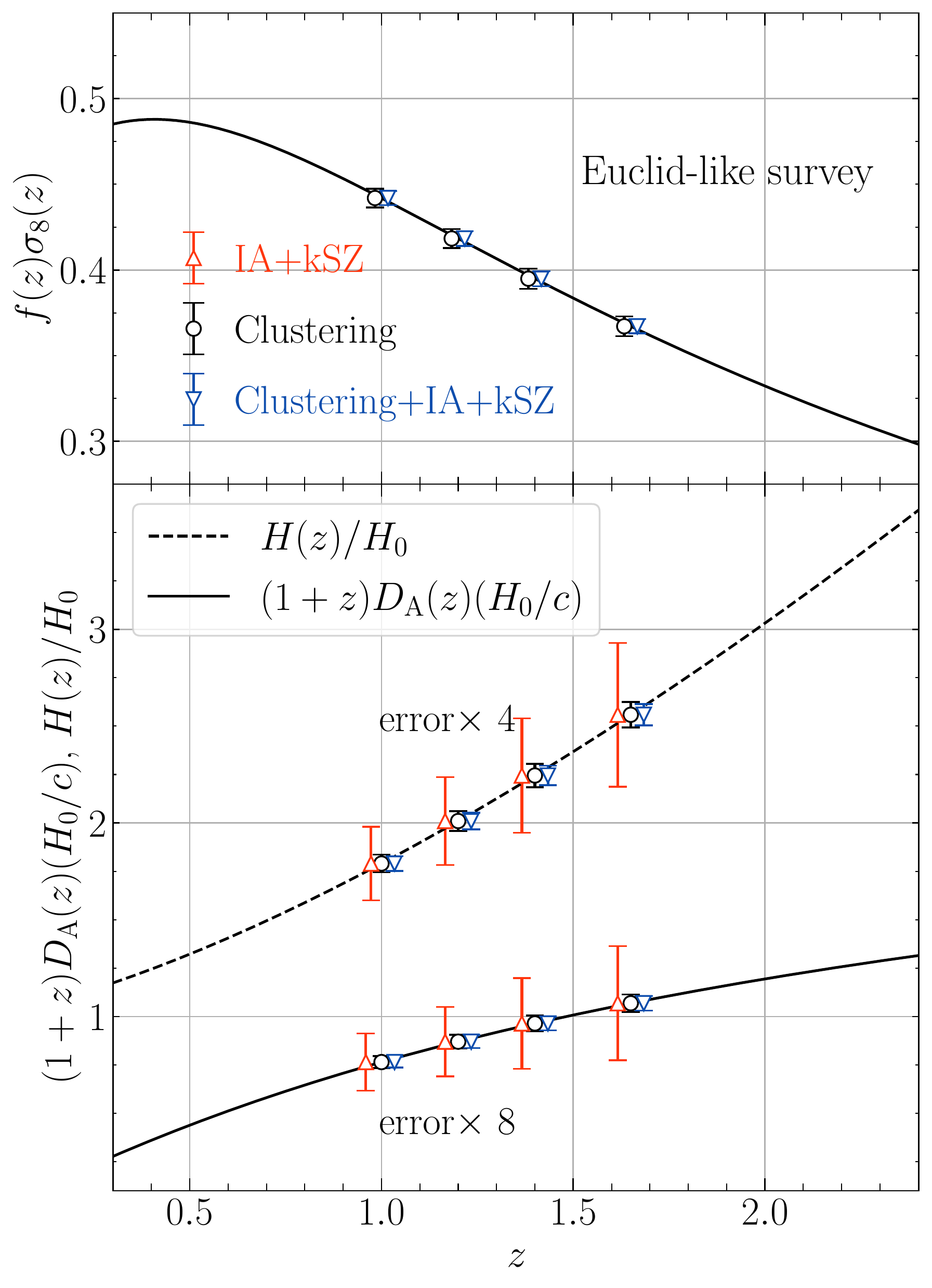}
\caption{ {\it Left set:} one-dimensional marginalized errors on the
  growth rate $f(z)\sigma_8(z)$ (upper panel) and geometric distances
  (lower panel), $D_A(z)$ and $H(z)$, expected from the deep
  (PFS-like) survey. {\it Right set:} same as the left set but the
  result expected from wide ({\it Euclid}-like) survey.  The errors on
  $H$ and $\DA$ are multiplied by 4 and 8, respectively, for
  illustration.  }
\label{fig:Errors_fsigma8_dA_H}
\end{figure*}

\subsection{Geometric and dynamical constraints}\label{sec:constraints1}

Let us first look at model-independent dynamical and geometric
constraints, namely the constraints on $f(z)\sigma_8(z)$, $\DA(z)$ and
$H(z)$, expected from the upcoming Subaru PFS survey.  From the
original Fisher matrix which includes these parameters in addition to
nuisance parameters, as summarized in table \ref{tab:statistics}, we
obtain the marginalized constraints as described in section
\ref{sec:fisher}.

The left panel of figure \ref{fig:2D_errors_PFS_f_dA_H} shows the
two-dimensional $1\sigma$ error contours on $f\sigma_8$, $\DA$ and $H$
normalized by their fiducial values, which are obtained individually
from galaxy clustering ($\Pgg$), kSZ ($\Pvv$) and IA ($\PEE$).  Since
the PFS is a deep survey and has seven redshift bins at $0.6<z<2.4$
(see Table \ref{tab:survey_pfs}), we here plot the result for the
central redshift bin, $1.2<z<1.4$, where the number density of
galaxies is the largest. Note that the left panel of figure
\ref{fig:2D_errors_PFS_f_dA_H} does not consider any cross correlation
between different probes, namely $\PgE$, $\Pgv$ and $\PvE$ (see table
\ref{tab:statistics}).  As clearly shown in the figure, using either
$\PEE$ or $\Pvv$ cannot constrain the growth rate. This is because the
intrinsic galaxy shapes themselves are insensitive to RSD in linear
theory and the kSZ only constrains the combination of $f\sigma_8$ and
$\tau$ without imposing any prior on $\tau$.  Nevertheless, each
single measurement of kSZ and IA can give meaningful constraints on
$\DA$ and $H$.  Then, including the cross correlation, the combination
of the two probes, namely $\Pvv$ and $\PEE$ as well as $\PvE$,
improves the constraint on $(\DA, H)$, depicted as the blue contour.

Interestingly, the constraining power on $\DA$ and $H$, when combining
kSZ and IA, can become tighter, and for the one-dimensional
marginalized error, the precision on each parameter achieves a few
percent level. Although the galaxy clustering still outperforms the
kSZ and IA observations, systematic effects in each probe come to play
differently (e.g., galaxy bias, shape noises and optical depth), and
in this respect, the geometric constraints from the kSZ and IA are
complementary as alternatives to those from the galaxy clustering.
Thus, constraining the geometric distances with kSZ and/or IA effects
would help addressing recent systematics-related issues such as the
Hubble tension.

The right panel of figure \ref{fig:2D_errors_PFS_f_dA_H} shows the
result similar to the left panel, but the joint constraints combining
kSZ and/or IA with galaxy clustering.  Compared to the results from
the single probe, the constraints are indeed improved, as previously
demonstrated in Refs. \cite{Sugiyama:2017} (clustering+kSZ) and
\cite{Taruya:2020} (clustering+IA).  Here we newly show that the
combination of all three probes, characterized by the six power
spectra, can further tighten the constraints on both the geometric
distances and growth of structure.  The results imply that adding any
of these power spectra can extract independent cosmological
information even though they measure the same underlying matter field.
The left panel of figure \ref{fig:Errors_fsigma8_dA_H} summarizes the
one-dimensional marginalized errors on $f\sigma_8$, $\DA$ and $H$
expected from the deep (PFS-like) survey, plotted as a function of $z$
over $0.6<z<2.4$.  Over all redshifts studied here, adding the
information from kSZ and IA measurements does improve the geometric
and dynamical constraints.

\begin{table*}[bt]
\caption{Summary of the cosmological models investigated in the forecast study.}
\begin{center}
\begin{tabular}{ l c l c c c c  }
\hline \hline 
Model            & No. of free & \multicolumn{3}{c}{Parameters $\{q_n\}$} & CMB prior & Result \\
\cline{3-5}
  &      parameters $N_q$                & Flat & Non-flat & MG     &     & (Fig.)         \\
\hline 
$w_0$ flat                            & 4 &$\Omegam,H_0,w_0,\sigma_8$        & $-$      & $-$                      & $-$    &\ref{fig:w0cdm}     \\
$w_0w_a$ flat                     & 5 &$\Omegam,H_0,w_0,w_a,\sigma_8$ & $-$      & $-$                      & Yes   &\ref{fig:w0wacdm}  \\
$w_0w_a$ non-flat              & 6 &$\Omegam,H_0,w_0,w_a,\sigma_8$ & $\OmegaK$ & $-$             & Yes   &\ref{fig:ow0wacdm}\\
$w_0w_a\gamma$ flat        & 6 &$\Omegam,H_0,w_0,w_a,\sigma_8$ & $-$      & $\gamma$          & Yes   &\ref{fig:gw0wacdm} \\
$w_0w_a\gamma$ non-flat & 7 &$\Omegam,H_0,w_0,w_a,\sigma_8$ & $\OmegaK$ & $\gamma$ & Yes   &\ref{fig:ogw0wacdm} \\
\hline \hline
\end{tabular}
\label{tab:models}
\end{center}
\end{table*}
%

\subsection{Cosmological parameter constraints}\label{sec:constraints2}

%
%
\begin{figure}[t]
\begin{center}
\includegraphics[width=0.9\columnwidth]{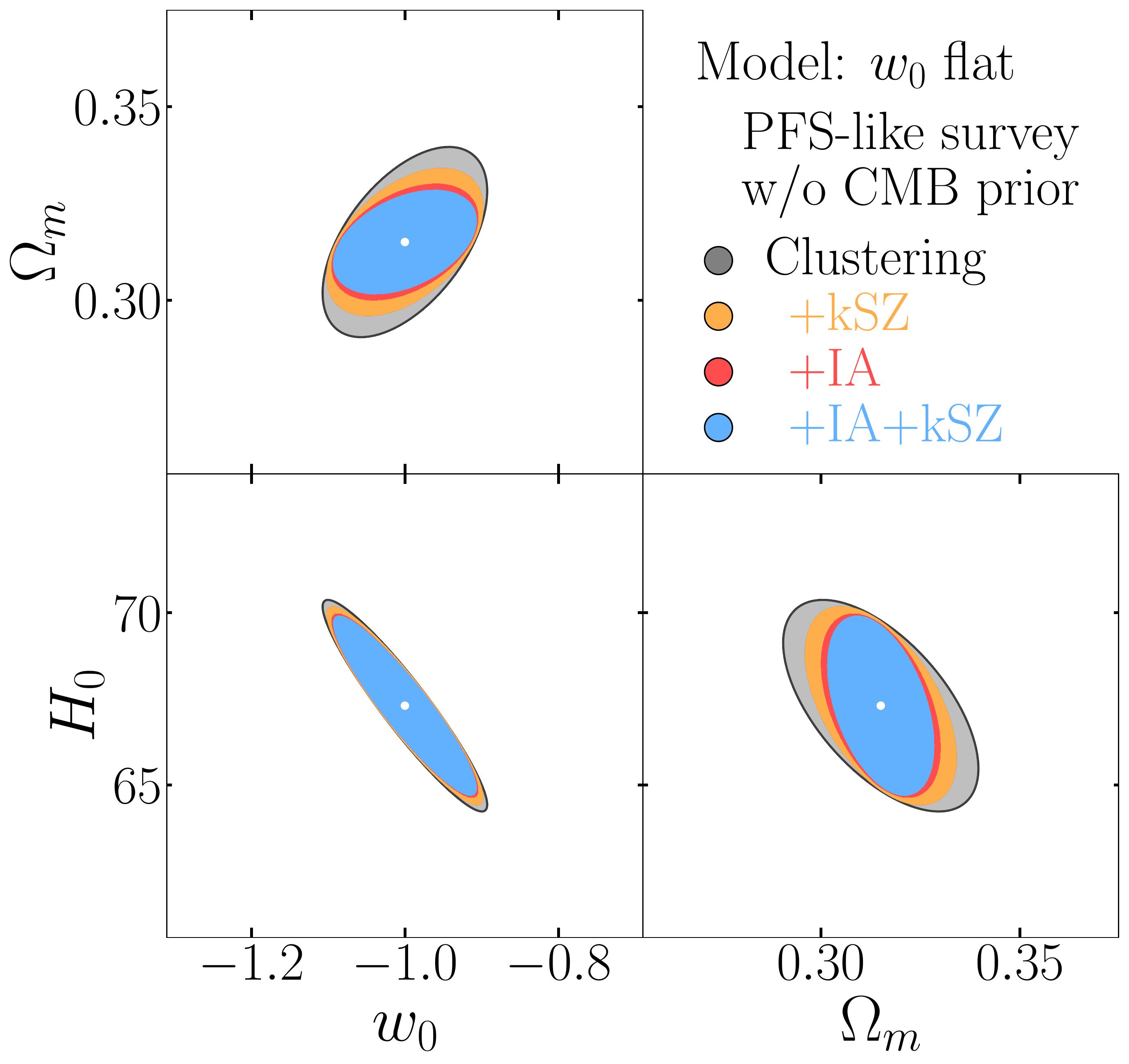}
\caption{Cosmological constraints on the $w_0$ flat model expected
  from the deep (PFS-like) survey without relying on the CMB prior. In
  each panel, contours show the $1\sigma$ confidence levels, with the
  amplitude parameter today, $\sigma_{8,0}=\sigma_8(0)$, marginalized
  over. Note that the joint constraints obtained from clustering and
  IA (red contours) are almost entirely behind those from clustering,
  IA and kSZ (blue contours).  }
\label{fig:w0cdm}
\end{center}
\end{figure}

\begin{figure}[t]
\begin{center}
\includegraphics[width=0.9\columnwidth]{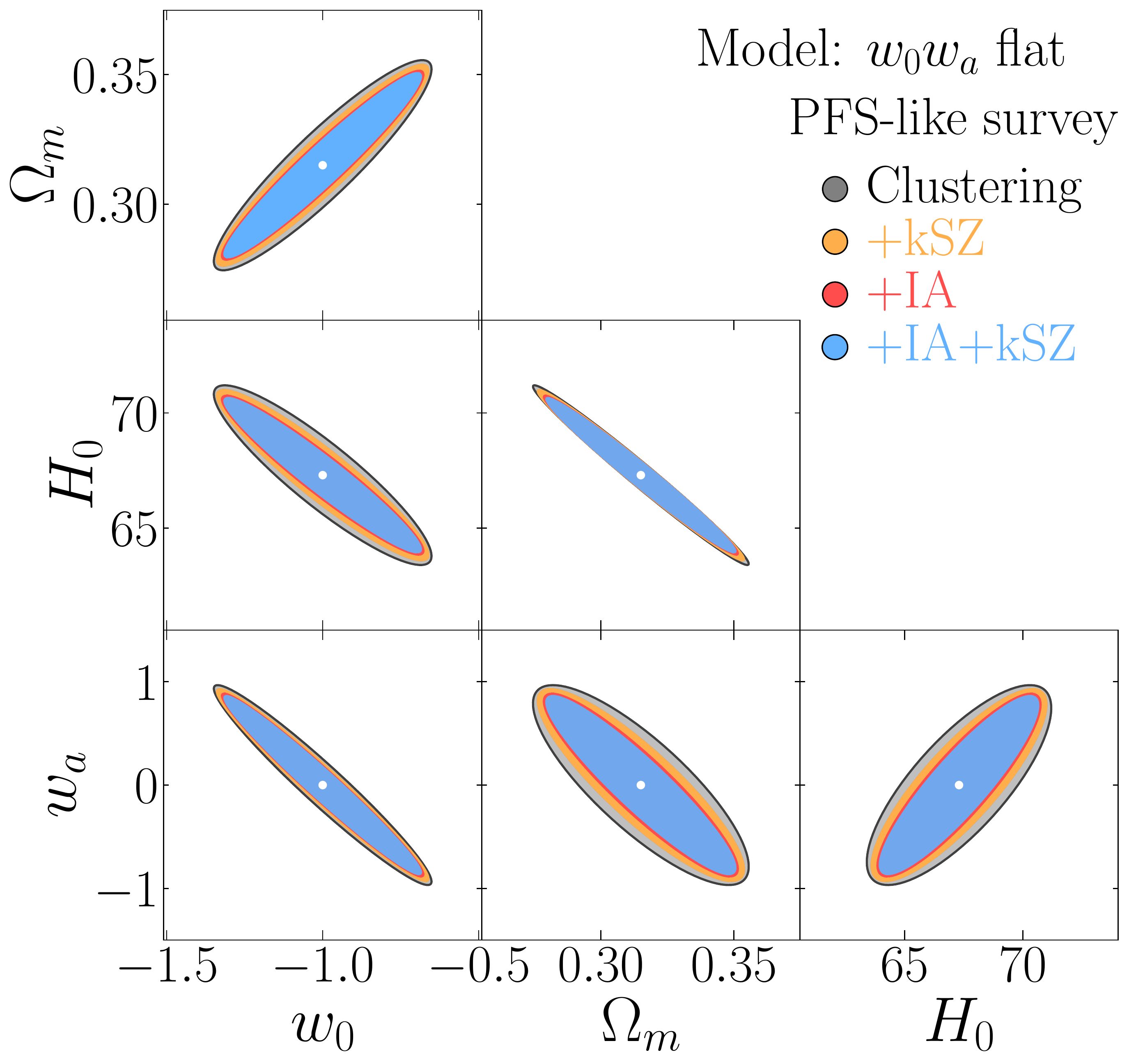}
\caption{Cosmological constraints on the $w_0w_a$ flat model expected
  from the deep (PFS-like) survey.  Unlike figure \ref{fig:w0cdm}, CMB
  prior information is added here. In each panel, contours show the
  $1\sigma$ confidence levels, with the amplitude parameter today,
  $\sigma_{8,0}=\sigma_8(0)$, marginalized over.  The joint
  constraints obtained from clustering and IA (red contours) are
  almost entirely behind those from clustering, IA and kSZ (blue
  contours).  }
\label{fig:w0wacdm}
\end{center}
\end{figure}

Provided the model-independent geometric and dynamical constraints
estimated from the original Fisher matrix in section
\ref{sec:constraints1}, we further discuss specific cosmological model
constraints listed in Table \ref{tab:models}.  In what follows, except
the $w_0$ flat CDM model, we add the CMB prior information to
constrain cosmological parameters and follow the conventional approach
adopted in the data analysis of BOSS
\cite{Eisenstein:2005,Percival:2007,Beutler:2017}, which do not use
the information of the full-shape power spectra. To be precise, we
introduce the following scaling parameters:
\be
\alpha_\parallel = \frac{H (z)r_d}{H^\fid(z)r_d^\fid}, \ \ \ \ \ \ 
\alpha_\perp     = \frac{\DA(z) r_d^\fid}{D_{\rm A}^{\fid}(z)r_d},
\ee
where the quantity $r_d$ is the sound horizon scale at the drag epoch
$z_d$ when photons and baryons are decoupled \cite{Eisenstein:1998}, 
given by
\be
r_d = \int^\infty_{z_d} \frac{c_s(z)}{H(z)}dz,
\ee
with $c_s$ being the sound speed in the photon-baryon fluid. We then
redefine the fiducial wavenumbers $k_\parallel$ and $k_\perp$, which
appear in Eq.~(\ref{eq:Pij_AP}), as $k_\parallel^\fid=k_\parallel
/\alpha_\parallel$ and $k_\perp^\fid=k_\perp\alpha_\perp$. With this
parameterization, the original expression for the power spectrum at
Eq.~(\ref{eq:Pij_AP}), taking the AP effect into account, is recast as
\be
P_{ij}^{\obs}\left(k_\perp^\fid,k_\parallel^{\fid};z \right) = 
\left(\frac{r_d^\fid}{r_d}\right)^3\frac{\alpha_\parallel}{\alpha_\perp^2}
P_{ij}\left(k_\perp,k_\parallel ; z \right), \label{eq:Pij_AP2}
\ee
where the prefactor
$(r_d^\fid/r_d)^3(\alpha_\parallel/\alpha_\perp^2)$ is equivalent to
that in equation (\ref{eq:Pij_AP}). Note that the dimensionless
quantities $r_d/D_{\rm A}$ and $H r_d$ are related to the actual BAO
scales measurable from galaxy surveys, i.e., angular separation and
redshift width of the acoustic scales. In this respect, with the form
given in equation (\ref{eq:Pij_AP2}), we are assuming that the main
contribution to the AP effect comes from the BAO.  As discussed in
Ref. \cite{Beutler:2017}, the uncertainty on the $r_d$ measurement
from the {\it Planck} experiment is only at the level of $\sim 0.2$
per cent \cite{Planck-Collaboration:2016} and fixing $r_d$ in equation
(\ref{eq:Pij_AP2}) has a negligible effect on our cosmological
parameter estimation. Based on this argument, we approximately set the
pre-factor $(r_d^\fid/r_d)^3$ to unity for the Fisher matrix analysis
below.


\begin{figure}[t]
\begin{center}
\includegraphics[width=0.995\columnwidth]{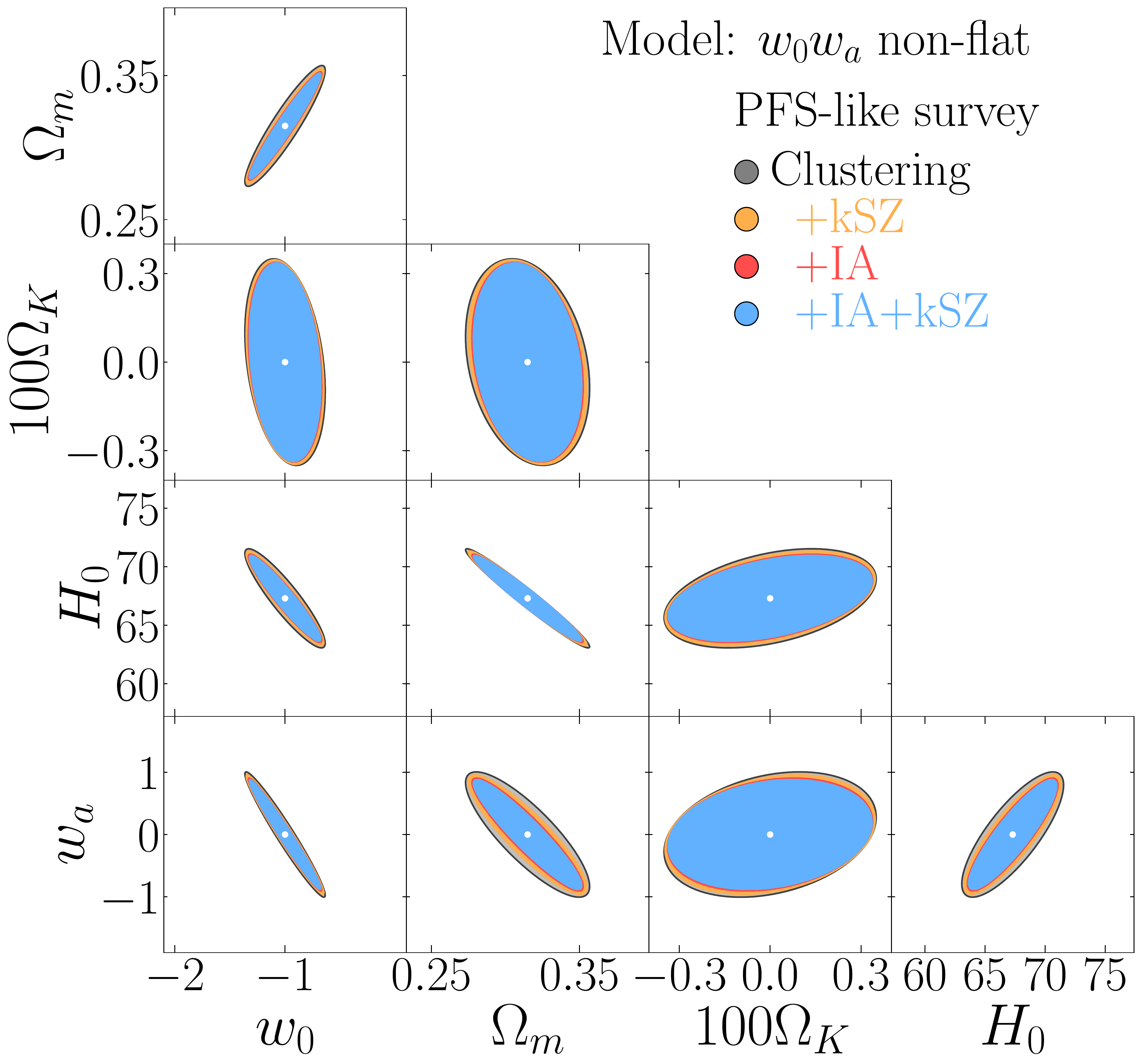}
\caption{
Same as figure \ref{fig:w0wacdm} but for the $w_0w_a$ non-flat model.
}
\label{fig:ow0wacdm}
\end{center}
\end{figure}


\begin{figure}[t]
\begin{center}
\includegraphics[width=0.98\columnwidth]{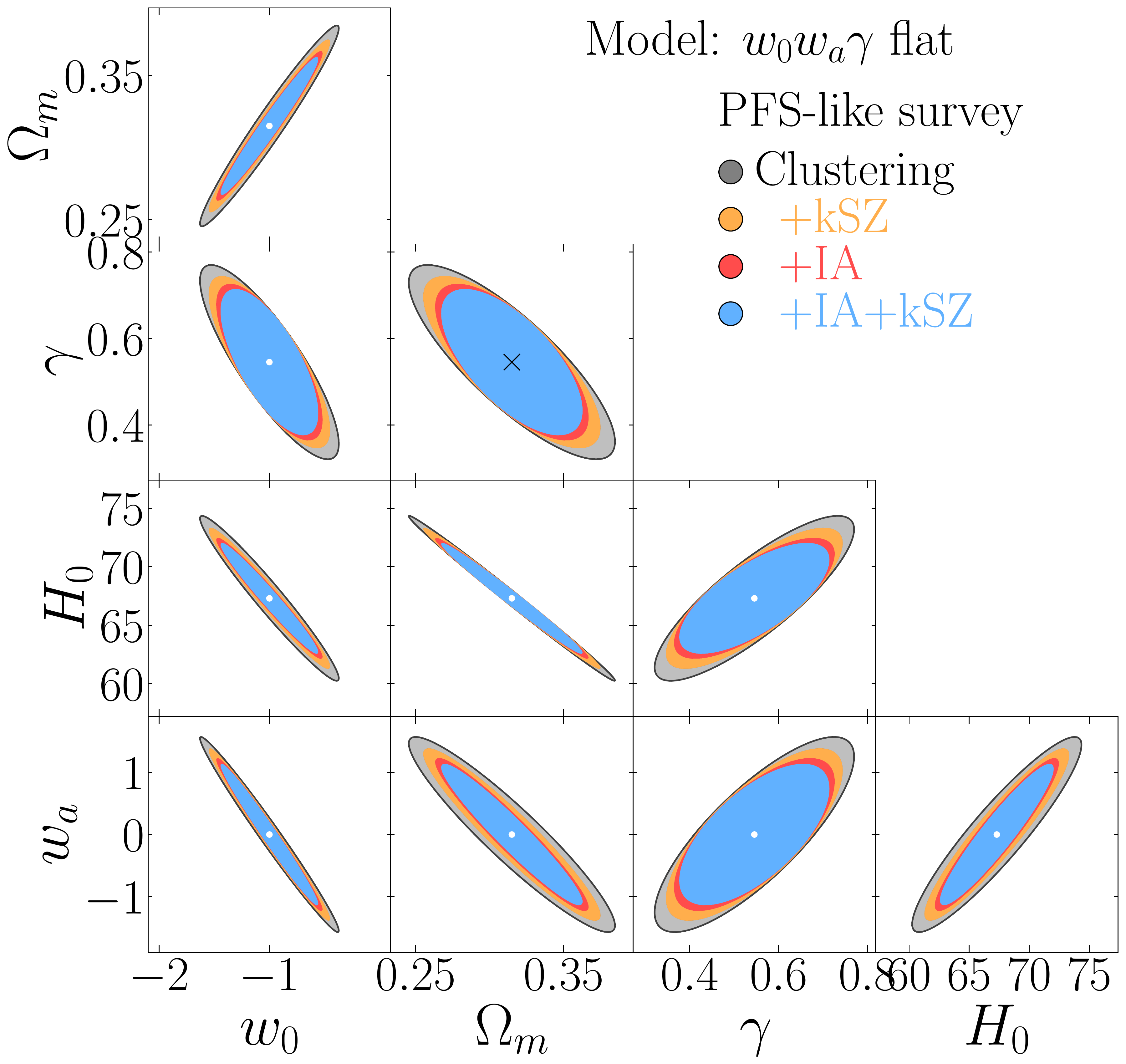}
\caption{
Same as figure \ref{fig:w0wacdm} but for the $w_0w_a\gamma$ flat model.
}
\label{fig:gw0wacdm}
\end{center}
\end{figure}


\begin{figure}[t]
\centering
\includegraphics[width=0.496\textwidth]{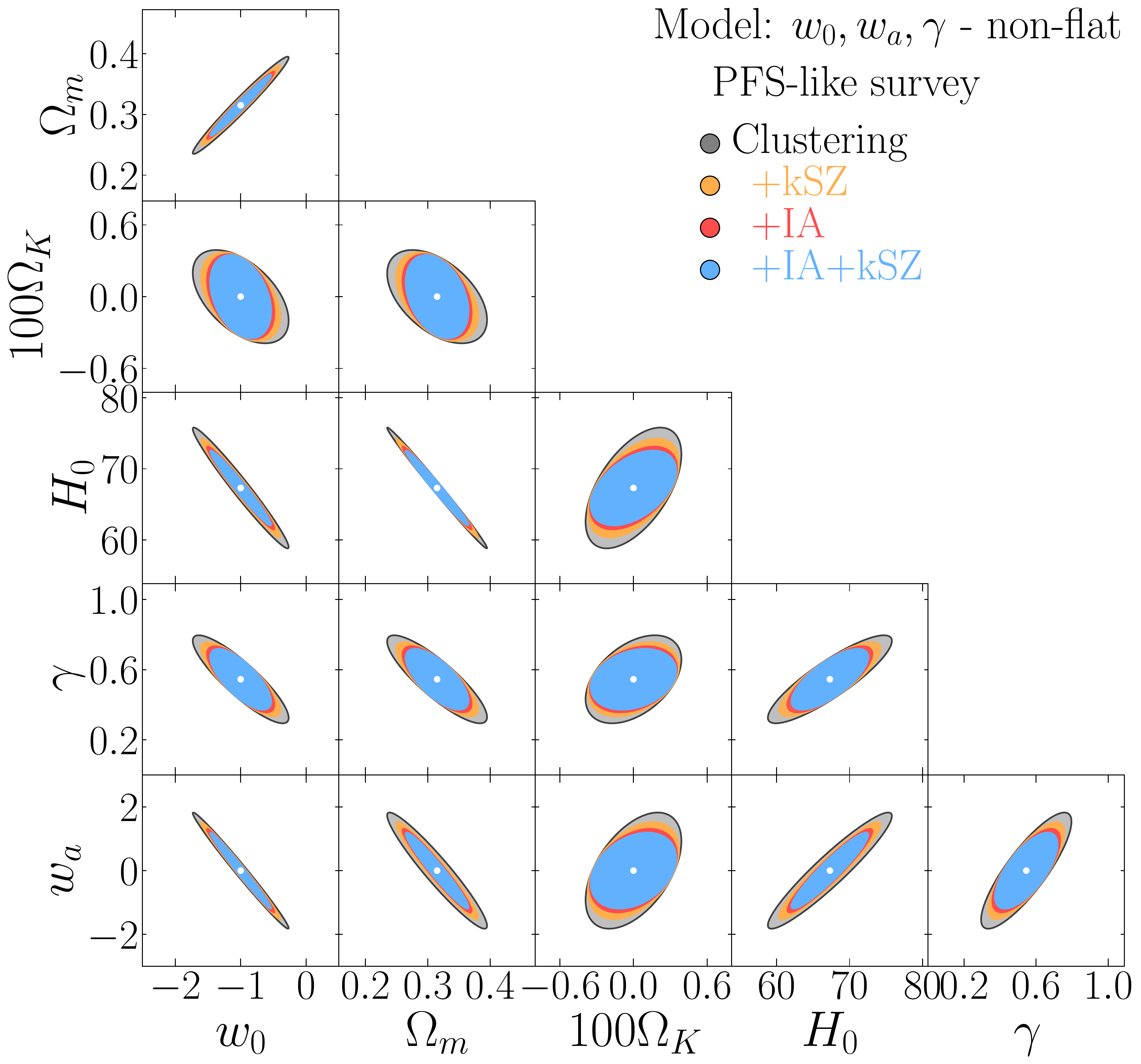}
\includegraphics[width=0.496\textwidth]{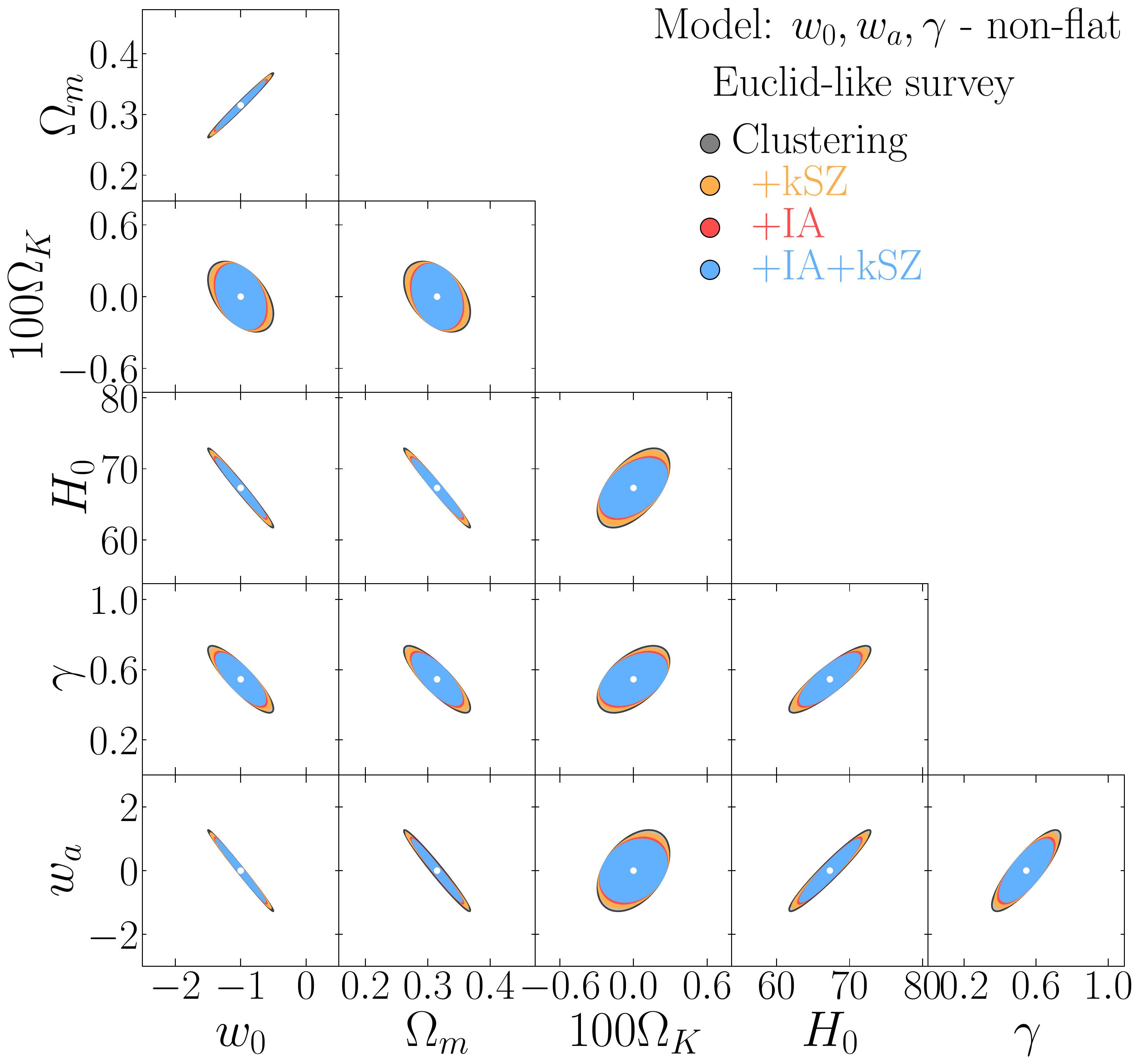}
\caption{ {\it Top}: Same as figure \ref{fig:w0wacdm} but for the
  $w_0w_a\gamma$ non-flat model from the deep (PFS-like) survey. {\it
    Bottom}: Similar to the top panel but from the wide ({\it
    Euclid}-like) survey.  }
\label{fig:ogw0wacdm}
\end{figure}

\begin{table*}[bt!]
\caption{Fractional marginalized errors on cosmological parameters,
  $\sigma/\theta^\fid$, for the four specific models. The CMB prior
  information is added for all the results here.  Since the fiducial
  values of $w_a$ and $\OmegaK$ are zero, we show the absolute errors,
  $\sigma$. Since the absolute errors on $\OmegaK$ are small, we show
  the errors multiplied by 100.}
\begin{center}
\begin{tabular}{l c c c c c c c c c c c  }
\hline \hline 
&& \ \ \  & \multicolumn{4}{c}{Deep (PFS-like) survey} &\ \ \ \ & \multicolumn{4}{c}{Wide ({\it Euclid}-like) survey}  \\
&&& Clustering &  &  & &&  Clustering &  &  & \\
Model & $\sigma/\theta^\fid$ & & only       & +kSZ         & +IA            & +IA+kSZ &&    only   & +kSZ        & +IA            & +IA+kSZ \\
\hline
$w_0,w_a$ & $\Omegam$ && 0.0850 & 0.0813 & 0.0765 & 0.0746 & & 0.0590 & 0.0578 & 0.0559 & 0.0555 \\
flat&$w_0$ && 0.230 & 0.224 & 0.213 & 0.208  && 0.164 & 0.163 & 0.157 & 0.156  \\
&$w_a$ && 0.638 & 0.613 & 0.584 & 0.569 && 0.455 & 0.450 & 0.438 & 0.434  \\
&$H_0$ && 0.0383 & 0.0363 & 0.0338 & 0.0329 && 0.0257 & 0.0251 & 0.0241 & 0.0239  \\
\hline
$w_0,w_a$ & $\Omegam$ &&  0.0877 & 0.0841 & 0.0788 & 0.0767  && 0.0592 & 0.0580 & 0.0561 & 0.0556  \\
non-flat & $w_0$ && 0.240 & 0.232 & 0.220 & 0.214 && 0.164 & 0.163 & 0.157 & 0.156  \\
&$w_a$ && 0.665 & 0.634 & 0.600 & 0.583 && 0.457 & 0.452 & 0.440 & 0.437  \\
&$H_0$ && 0.0415 & 0.0397 & 0.0370 & 0.0360 && 0.0262 & 0.0256 & 0.0247 & 0.0244 \\
&$100\OmegaK$ && 0.231 & 0.229 & 0.223 & 0.222 && 0.162 & 0.161 & 0.161 & 0.161 \\
\hline
$w_0,w_a,\gamma$ &$\Omegam$ && 0.1459 & 0.1244 & 0.1075 & 0.0990 && 0.1004 & 0.0955 & 0.0840 & 0.0804  \\
flat & $w_0$ && 0.415 & 0.359 & 0.314 & 0.288 && 0.288 & 0.272 & 0.245 & 0.234  \\
&$w_a$ && 1.036 & 0.907 & 0.804 & 0.743 && 0.761 & 0.715 & 0.655 & 0.625  \\
&$H_0$ && 0.0691 & 0.0585 & 0.0501 & 0.0458 && 0.0452 & 0.0429 & 0.0374 & 0.0358  \\
&$\gamma$ && 0.271 & 0.238 & 0.217 & 0.202 && 0.193 & 0.182 & 0.169 & 0.161  \\
\hline
$w_0,w_a,\gamma$ &$\Omegam$ && 0.1679 & 0.1380 & 0.1162 & 0.1055 && 0.1114 & 0.1039 & 0.0884 & 0.0836  \\
non-flat& $w_0$ && 0.484 & 0.400 & 0.340 & 0.307 && 0.330 & 0.304 & 0.264 & 0.248  \\
&$w_a$ && 1.202 & 1.004 & 0.865 & 0.786 && 0.841 & 0.774 & 0.686 & 0.646 \\
&$H_0$ && 0.0833 & 0.0682 & 0.0571 & 0.0516 && 0.0548 & 0.0510 & 0.0430 & 0.0404 \\
&$100\OmegaK$ && 0.258 & 0.245 & 0.234 & 0.230 && 0.194 & 0.189 & 0.182 & 0.179 \\
&$\gamma$ && 0.304 & 0.256 & 0.228 & 0.210 && 0.231 & 0.212 & 0.191 & 0.179  \\
\hline
\end{tabular}
\label{tab:constraints}
\end{center}
\end{table*}

Now, the model-independent parameters in our original Fisher matrix,
combining all three probes, become $\theta_\alpha=(b\sigma_8,
\AIA\sigma_8,\tau, f\sigma_8,\alpha_\parallel,\alpha_\perp)$, and the
marginalized constraints on $\vartheta_A=(f\sigma_8, \alpha_\perp,
\alpha_\parallel)$ are evaluated for each $z$-slice by constructing
the $3\times 3$ sub-matrix $\overline{\bfF}_{\rm LSS}(z_k)$. Summing
up these sub-matrices over all the redshift bins, i.e.,
$\overline{\bfF}_{{\rm LSS}} = \sum_{k} \overline{\bfF}_{{\rm
    LSS}}(z_k) $, we project it into a new parameter space to test the
model-dependent cosmological parameters $q_n$ through equation
(\ref{eq:projection_fisher}). The most general model considered in our
analysis is the $w_0w_a\gamma$ non-flat model, with
$q_n=(\Omegam,w_0,w_a,H_0,\OmegaK,\gamma,\sigma_{8,0})$. All the
cosmological models we consider in this paper are summarized in table
\ref{tab:models}.

\begin{figure*}[bt]
\centering
\includegraphics[width=0.481\textwidth]{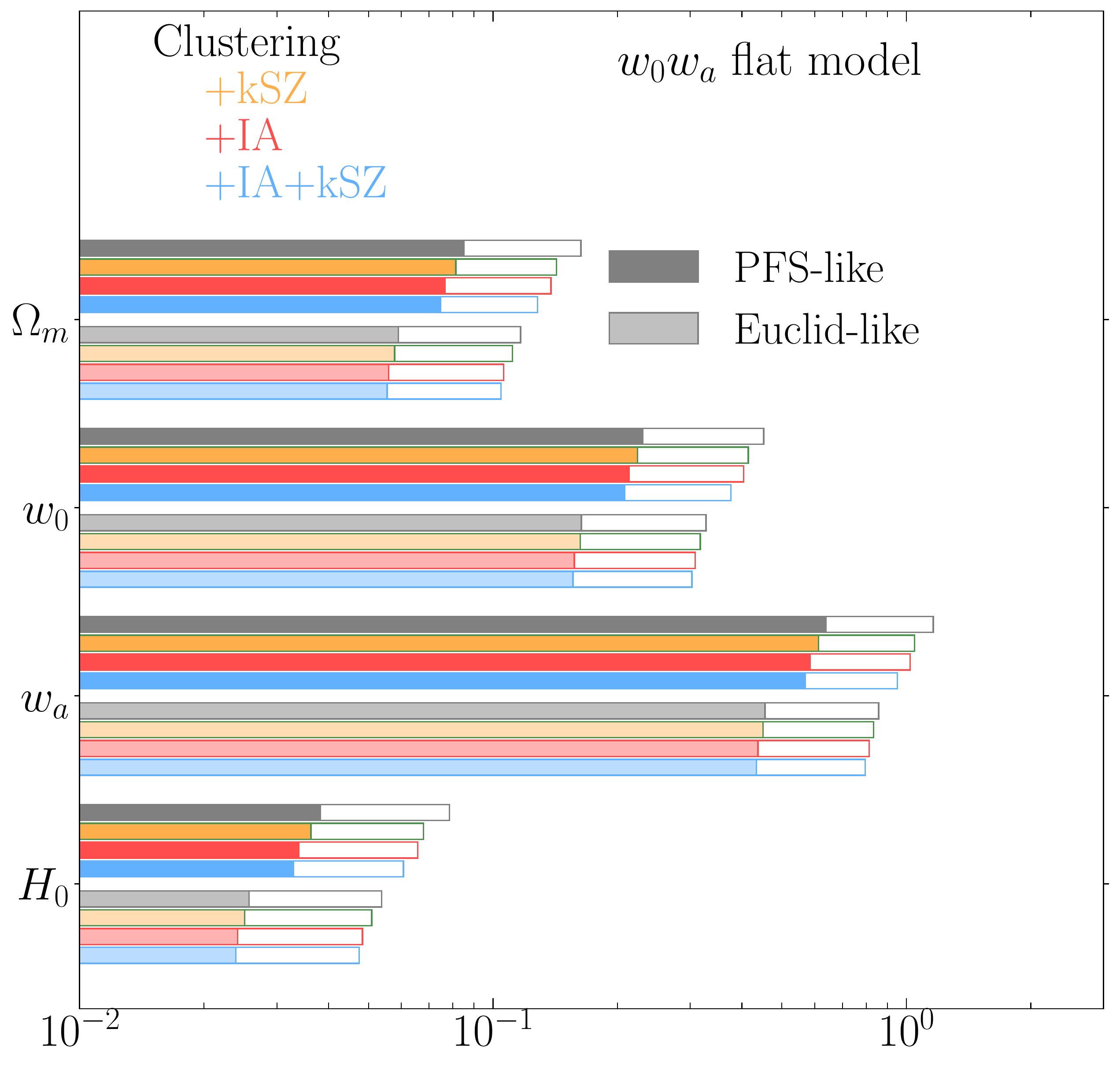}
\includegraphics[width=0.481\textwidth]{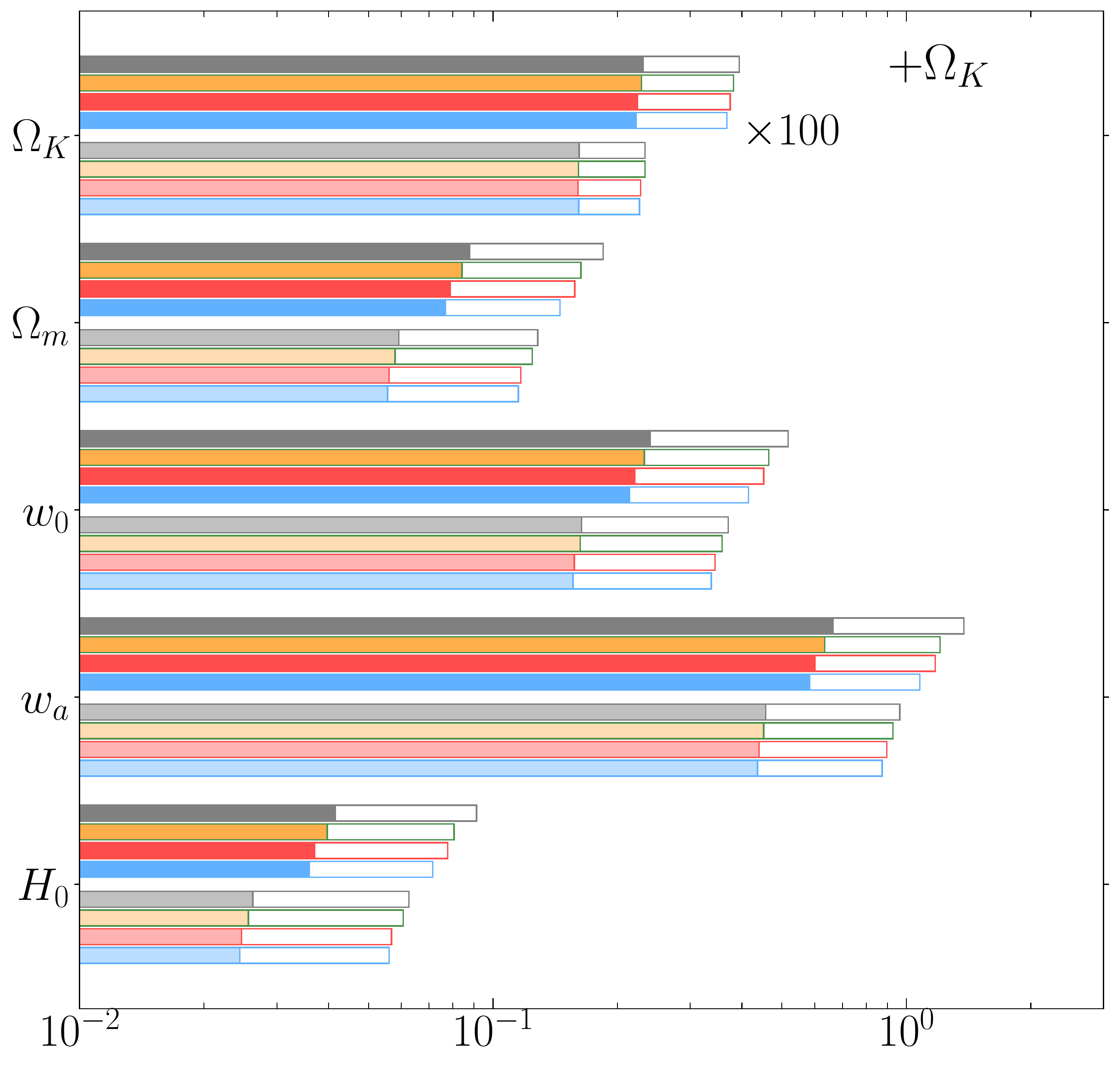}
\includegraphics[width=0.481\textwidth]{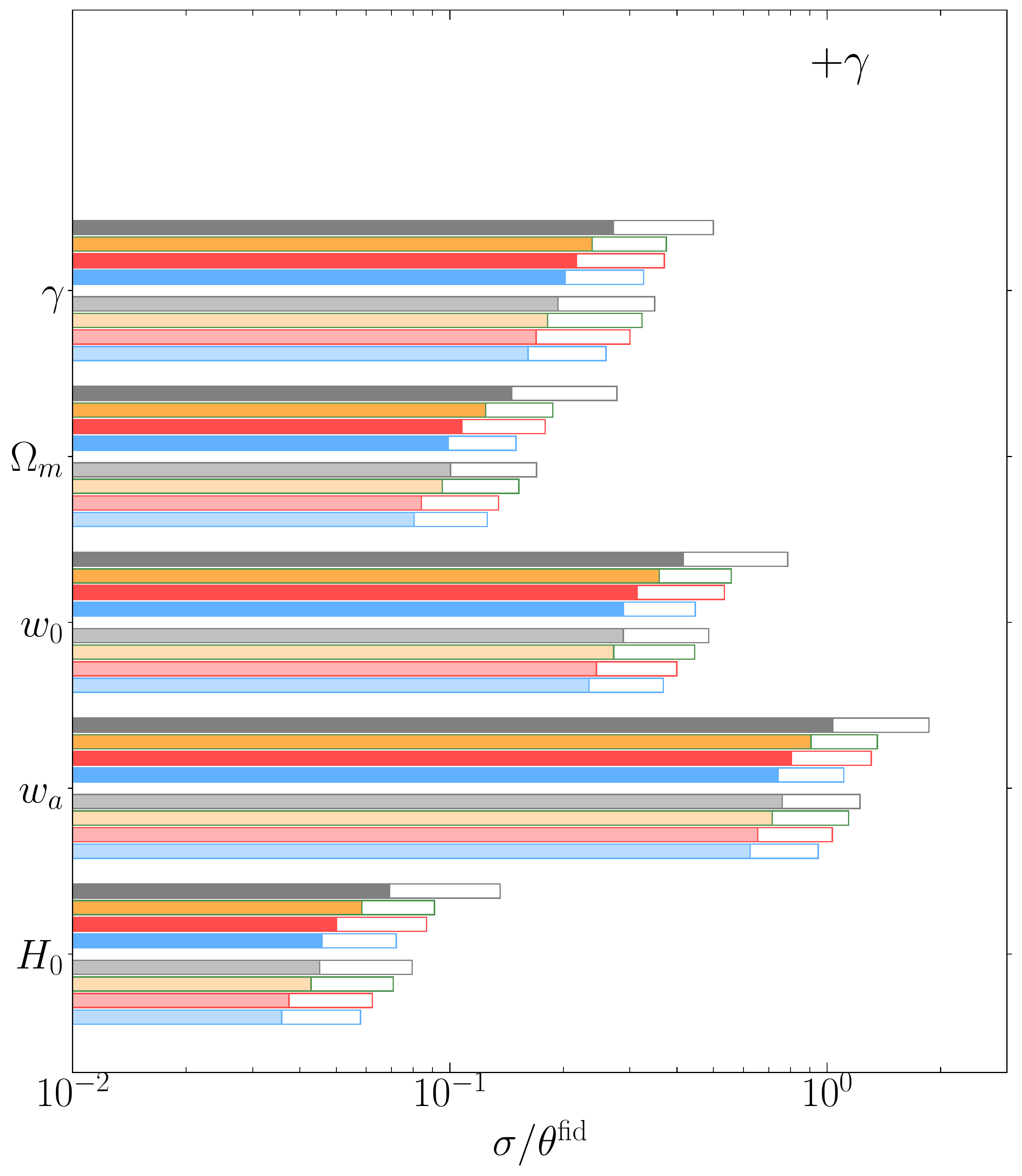}
\includegraphics[width=0.481\textwidth]{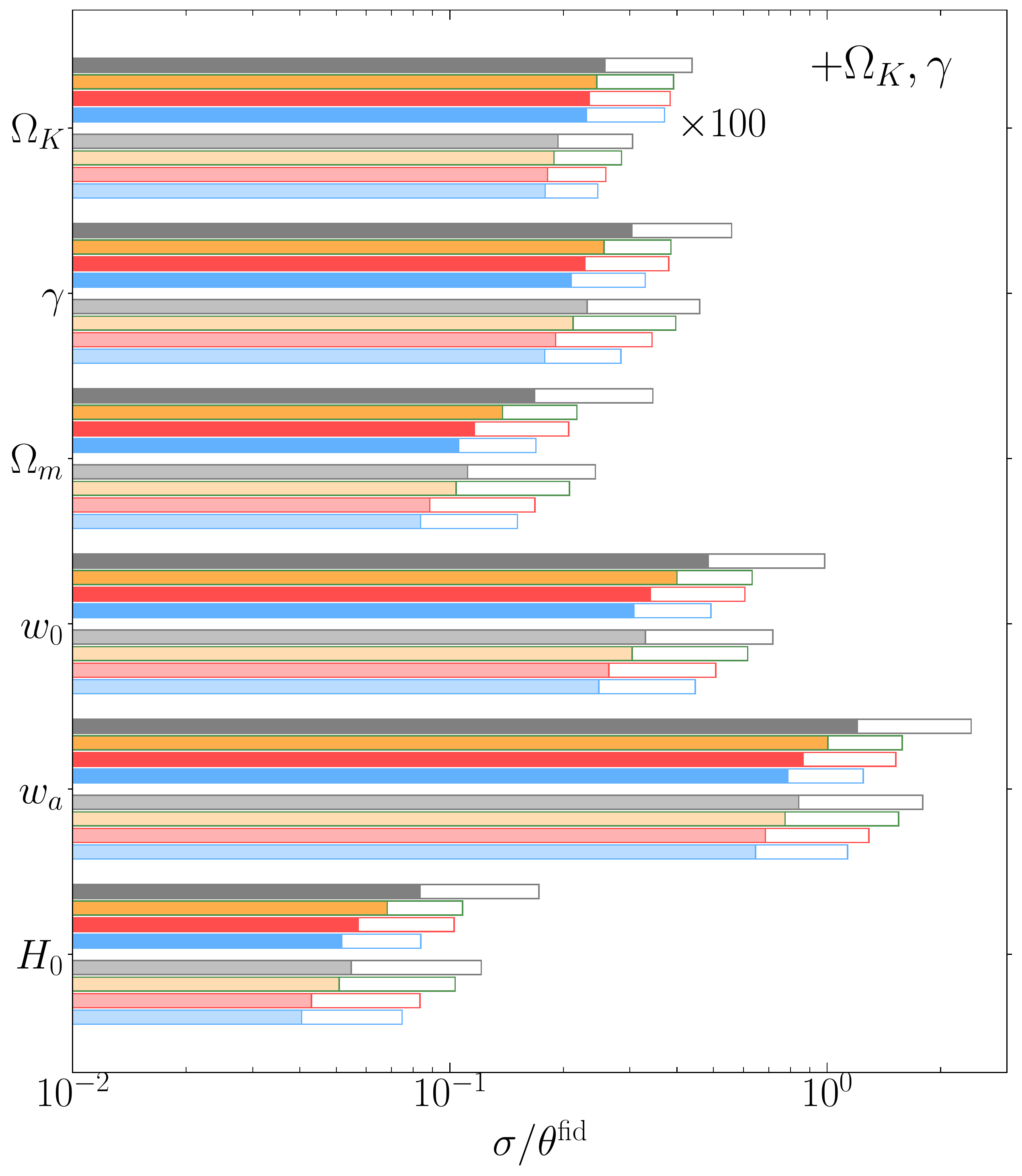}
\caption{Marginalized $1\sigma$ errors on cosmological parameters,
  relative to their corresponding fiducial values,
  $\sigma/\theta^\fid$.  The darkly and lightly filled bars show the
  errors from the deep (PFS-like) and wide ({\it Euclid}-like)
  surveys, respectively.  The top-left and top-right panels are for
  the $w_0w_a$ flat and non-flat models, respectively. Similarly, the
  bottom-left and bottom-right panels are for $w_0w_a\gamma$ flat and
  non-flat models, respectively. The CMB prior information is added
  for all the results here. Since the fiducial values of $w_a$ and
  $\OmegaK$ are zero, we show the absolute errors, $\sigma$.  Since
  the absolute errors on $\OmegaK$ are small, we show the errors
  multiplied by 100.  The hollow bars are similar with the filled bars
  but based on the conservative analysis with the scales of $k\leq
  0.1\,h\,{\rm Mpc}^{-1}$ (see appendix \ref{sec:conservative}).  }
\label{fig:cosmo_PFS_Euclid}
\end{figure*}

Let us show our main results for the deep, PFS-like survey below.
Figures \ref{fig:w0cdm} -- \ref{fig:gw0wacdm} and the top panel of
figure \ref{fig:ogw0wacdm} plot the expected two-dimensional
constraints on pairs of model parameters for different cosmological
models.  Also, table \ref{tab:constraints} and figure
\ref{fig:cosmo_PFS_Euclid} summarize the one-dimensional marginalized
constraints.  We will discuss all the results in detail in the rest of
this subsection.  Except for figure \ref{fig:w0cdm}, all the following
results are obtained adding the CMB prior information, as detailed in
Appendix \ref{sec:prior}.  Thus, the constraints are obtained from the
combination of the Fisher matrices of the LSS and CMB, $\bfS =
\bfS_{\rm LSS} + \bfS_{\rm CMB}$.  For all cases, the nuisance
parameters characterizing the power spectrum normalization on each
probe namely $b\sigma_8$, $\tau$, and $\AIA\sigma_8$, are marginalized
over. Comparisons of the obtained constraints with those from the
wide, {\it Euclid}-like survey will be presented in section
\ref{sec:deep_vs_wide}.

Figure \ref{fig:w0cdm} shows the case for the $w_0$ flat model, in
which we vary $q_n = (\Omega_m, w_0, H_0, \sigma_{8,0})$.  Only for
this model, we do not add the CMB prior and use LSS probes as our
primary data set.  As shown in Ref. \cite{Taruya:2020}, adding IA to
galaxy clustering significantly improves the constraints.  If the kSZ
measurement is added, one can achieve a similar (but slightly weaker)
improvement.  Simultaneously analyzing galaxy clustering with kSZ and
IA, the constraint on each cosmological parameter gets even tighter,
by $15-21\%$, compared to the clustering-only constraints.

In figure \ref{fig:w0wacdm}, adding the CMB prior information, we show
an extension of the parameter space by allowing the time-varying dark
energy equation-of-state, which is the $w_0w_a$ flat model described
by the parameters $q_n = (\Omegam, w_0, w_a, H_0, \sigma_{8,0})$.
Here, the improvement by adding IA is not so significant compared to
the former case, due mainly to a dominant contribution from the CMB
prior, consistent with the result of Ref.~ \cite{Taruya:2020}.
However, combining the galaxy clustering with both kSZ and IA
measurements, we can improve the constraints further, for example, on
$w_a$ by $\sim 11\%$, as shown in table \ref{tab:constraints} and
figure \ref{fig:cosmo_PFS_Euclid}.  Figure \ref{fig:ow0wacdm} examines
the case with non-zero $\OmegaK$, by introducing another degree of
freedom in the parameter space on top of the $w_0w_a$ flat model.
Note that based on the BAO experiments at high $z$, a best achievable
precision on $\OmegaK$, limited by the cosmic variance, has been
studied in detail in Ref.~\cite{Takada:2015}.  In our forecast, the
spatial curvature has already been tightly constrained by the CMB
prior.  Thus, the resulting constraints are similar with those of the
model with $\OmegaK=0$ in figure \ref{fig:w0wacdm}.


\begin{figure*}[t]
\centering
\includegraphics[width=0.49\textwidth]{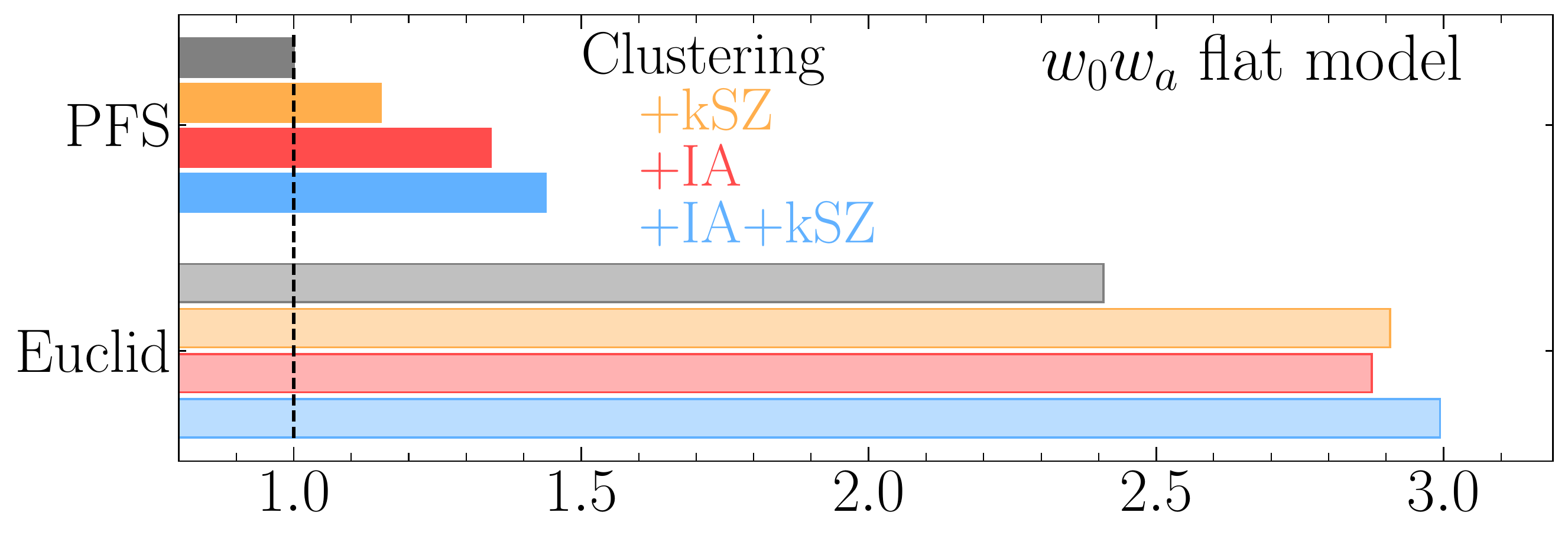}
\includegraphics[width=0.49\textwidth]{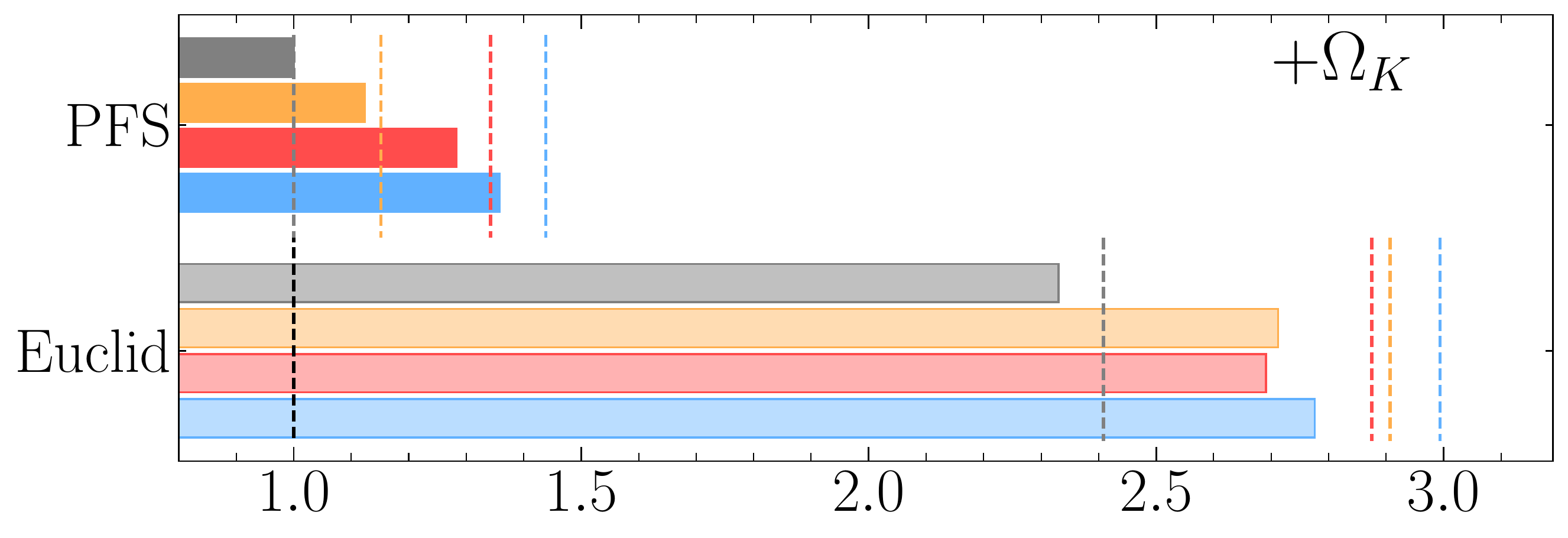}
\includegraphics[width=0.49\textwidth]{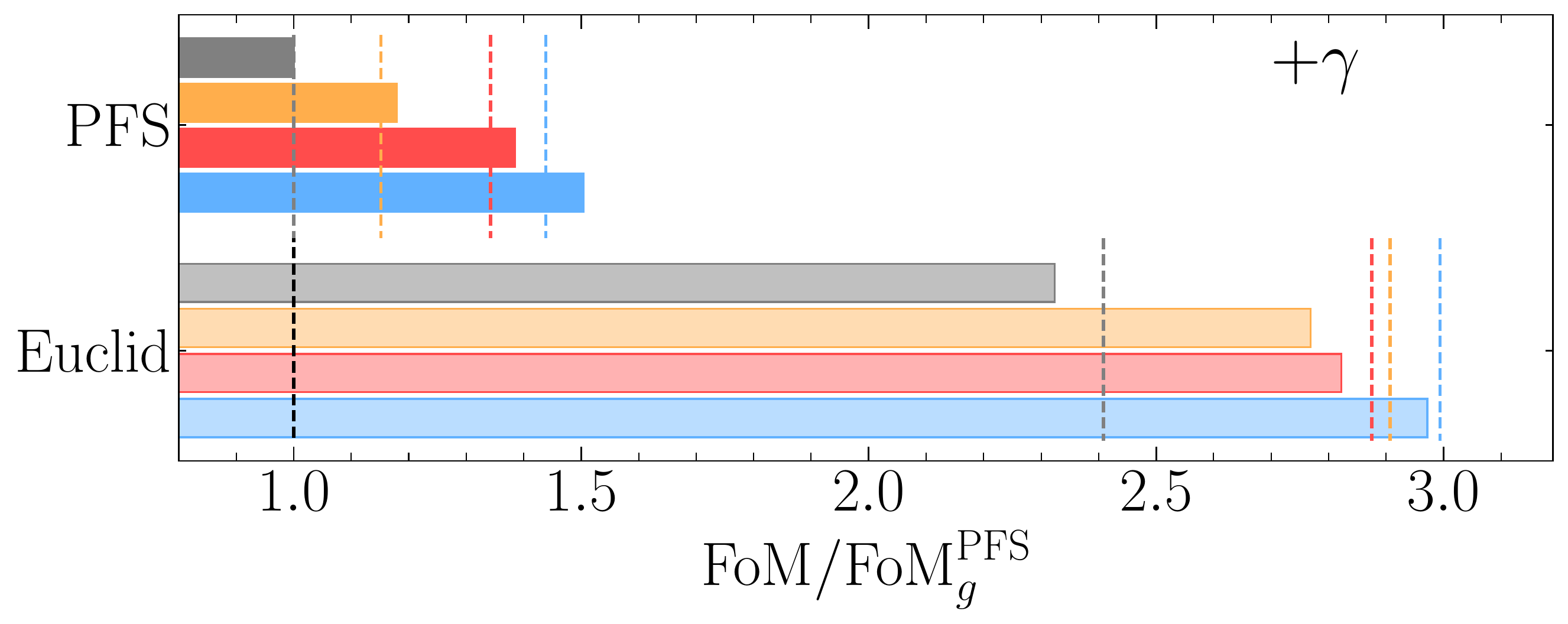}
\includegraphics[width=0.49\textwidth]{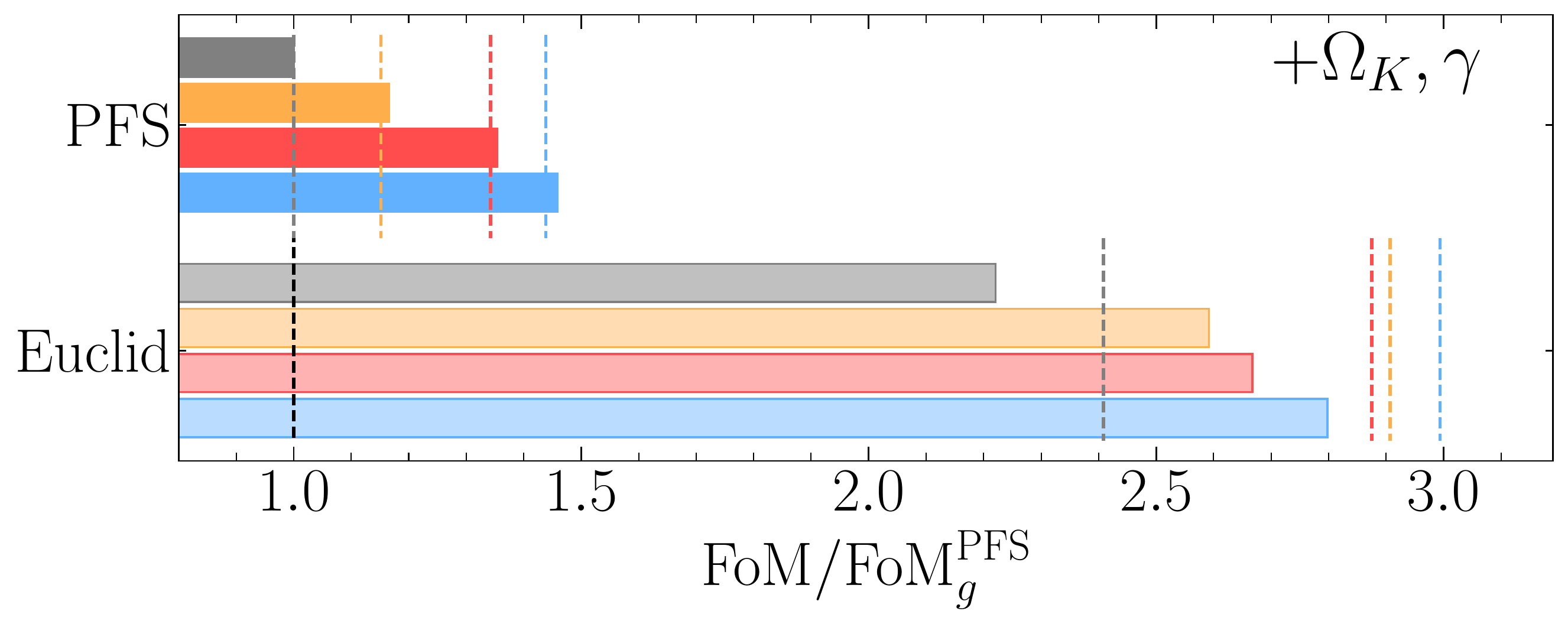}
\caption{ FoM of cosmological impact for clustering only (gray),
  clustering $+$ kSZ (yellow), clustering $+$ IA (red) and clustering
  $+$ IA $+$ kSZ (blue). The upper-left and upper-right panels are
  the results for the $w_0w_a$ flat and non-flat models,
  respectively. Similarly, the lower-left and lower-right panels are
  the results for the $w_0w_a\gamma$ flat and non-flat models,
  respectively. The CMB prior information is added for all the
  results here. In each panel, the values of FoM are normalized by
  that for the PFS survey with clustering-only analysis.  The yellow,
  red and blue vertical lines indicate the FoM values obtained in the
  upper-left panel for comparison.  }
\label{fig:FoM_PFS_Euclid}
\end{figure*}

\begin{figure*}[bt]
\centering
\includegraphics[width=0.49\textwidth]{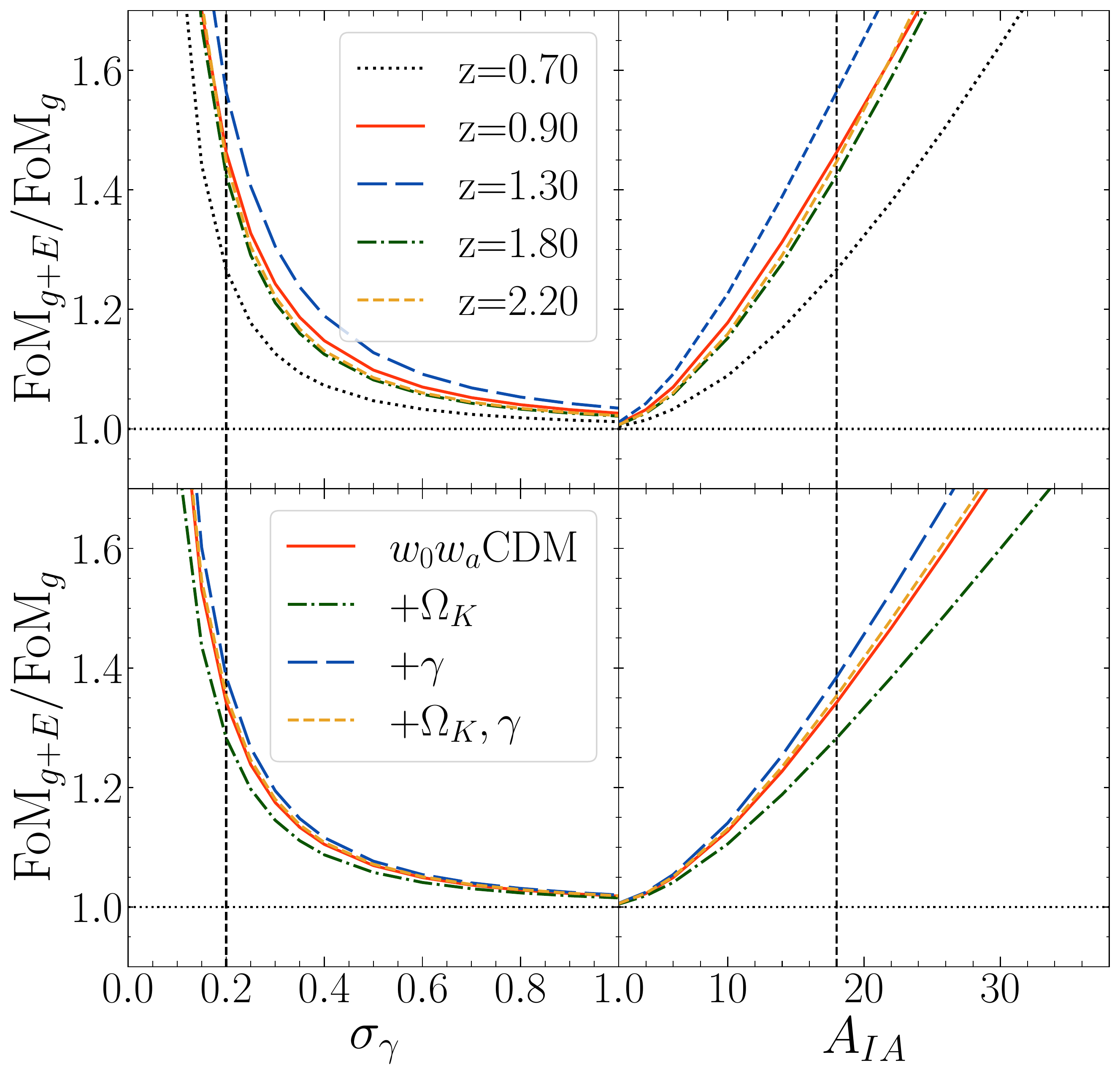}
\includegraphics[width=0.49\textwidth]{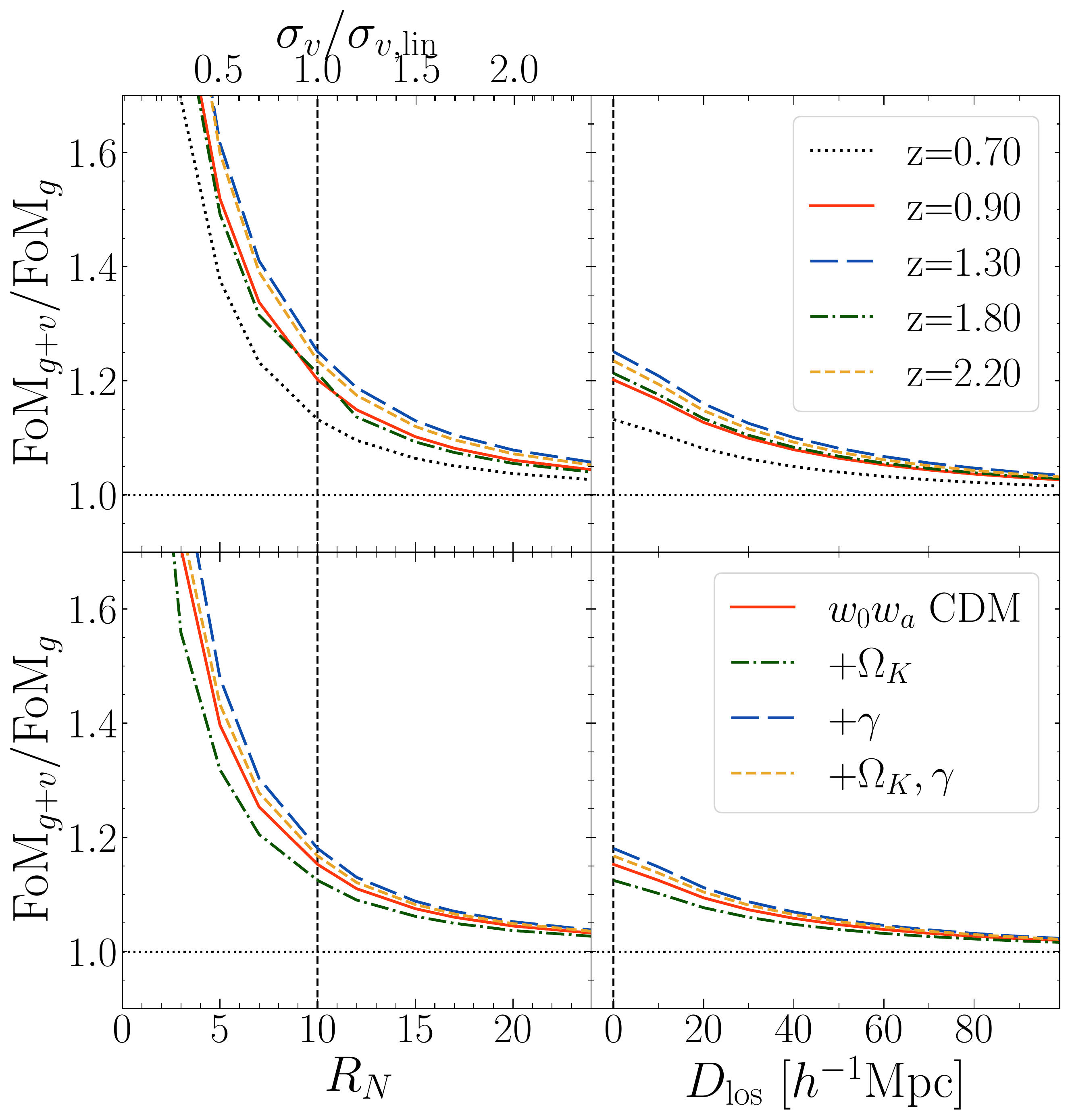}
\caption{Relative impact of combining IA ({\it left set}) and kSZ
  ({\it right set}) on the parameter constraints, defined by the ratio
  of figure-of-merit, $\FoM_{g+E}/\FoM_{g}$ and $\FoM_{g+v}/\FoM_{g}$,
  respectively. The subscripts of $g$, $g+E$ and $g+v$ denote the FoM
  expected from galaxy clustering only, the combination of clustering
  and IA, and that of clustering and kSZ, respectively. Upper panels
  show the results for geometric distances and structure growth,
  $\DA$, $H$, $f\sigma_8$, derived from each redshift slice of the PFS
  survey.  Results at $z=1.1$ and $z=1.5$ are not shown here because
  they are almost equivalent to those at $z=0.9$.  Bottom panels are
  the results for cosmological parameters, with $\sigma_{8,0}$
  marginalized over. In all cases, the vertical dashed lines indicate
  the default parameter setup (see Sec.~\ref{sec:setup}). }
\label{fig:FoM_PFS}
\end{figure*}

Now, allowing the deviation of growth of structure from the GR
prediction, characterized by the parameter $\gamma$, we test and
constrain both the cosmic expansion and gravity law, shown in figure
\ref{fig:gw0wacdm} and the top panel of figure
\ref{fig:ogw0wacdm}. Figure \ref{fig:gw0wacdm} considers the
$w_0w_a\gamma$ flat model, in which the spatial curvature is kept
flat. The resulting constraints from the clustering-only analysis are
generally weaker than the case of $w_0w_a$ non-flat model despite the
fact that the number of parameters remains unchanged. The main reason
comes from the newly introduced parameter $\gamma$, which can be
constrained only through the measurement of the growth rate, and is
strongly degenerated with $\Omegam$.  Nevertheless, adding the
information from the observations of kSZ and/or IA, the constraints
get significantly tighter, and combining all three probes, the
achievable precision is improved by $25\%$ for $\gamma$, and $\sim
30\%$ for other parameters, as shown in table \ref{tab:constraints}
and figure \ref{fig:cosmo_PFS_Euclid}. In the top panel of figure
\ref{fig:ogw0wacdm}, the significance of combining all three probes is
further enhanced in $w_0w_a\gamma$ non-flat model, where we have seven
parameters of $q_n = (\Omegam, \OmegaK, w_0, w_a, H_0, \gamma,
\sigma_{8,0})$.  As a result, compared to the clustering-only
analysis, the simultaneous analysis with the clustering, IA and kSZ
further improves the constraints by $31\%$ for $\gamma$ and $>35\%$
for others except for $\OmegaK$ (see table \ref{tab:constraints} and
figure \ref{fig:cosmo_PFS_Euclid}).

\section{Discussion}\label{sec:discussion}

\subsection{Deep vs wide surveys}\label{sec:deep_vs_wide}

So far, we have considered the PFS survey as a representative example
of deep galaxy surveys. Here, we discuss how the constraining power of
kSZ and IA measurements depends on types of galaxy surveys.  For this
purpose, we perform the forecast analysis for the {\it Euclid} survey
as an example of wide galaxy survey.  The right panel of figure
\ref{fig:Errors_fsigma8_dA_H} presents geometric and dynamical
constraints from the {\it Euclid}-like survey. Though the redshift
range for the {\it Euclid} is narrower than that for the PFS, the
constraints on $f\sigma_8$, $\DA$ and $H$ at each redshift bin are
much tighter due to the large survey volumes (see table
\ref{tab:survey_euclid}).  Cosmological constraints are thus expected
to be stronger as well.  To see it quantitatively, let us utilize the
FoM introduced in equation (\ref{eq:FoM}).  Here, we marginalize over
the amplitude parameter today, $\sigma_{8,0}$, via the inversion of
the $N_q \times N_q$ Fisher matrix, $\bfS$ (see equation
(\ref{eq:projection_fisher})). The size of the matrix
$\overline{\bfS}$ is thus ($N_q-1)\times(N_q-1)$.  Indeed, the FoM for
cosmological parameters from the wide survey is always better, roughly
by a factor of two, than that from the PFS.  The comparison is shown
for the four cosmological models in figure \ref{fig:FoM_PFS_Euclid}.

Constraints on each cosmological parameter is made with the projection
of the Fisher matrix. The forecast results from the {\it Euclid}
survey are summarized in the right hand side of table
\ref{tab:constraints} and figure \ref{fig:cosmo_PFS_Euclid}.  If one
uses only the information of clustering, constraints from the wide
survey considered here are always tighter than those from the deep
survey, by $25-40\%$.  Then one can improve the constraints by the
joint analysis of clustering, IA and kSZ, similarly to the analysis of
deep galaxy surveys.  However, the improvement of the cosmological
constraints are not so significant as the case of the deep survey. It
is particularly prominent if we consider the model which allows the
$\gamma$ parameter to vary.  For example, in the $w_0w_a\gamma$ flat
model, while the improvement of cosmological parameters for the deep
survey is $25-34\%$, that for the wide survey is $17-21\%$. It could
be due to the fact that the $\gamma$ parameter is constrained from the
redshift dependence of the measured growth rate $f(z)$ at various
redshifts, and thus the constraining power in the wide survey does not
gain as much as that in a deep survey by combining with additional
probes of kSZ and IA.  As a result, if we perform a joint analysis of
galaxy clustering together with kSZ and IA for a deep survey, the
constraining power can be as strong as the conventional
clustering-only analysis for a wide survey even though the FoM for the
wide survey is twice as large.  More interestingly, in the most
general $w_0w_a\gamma$ non-flat model, even the deep survey with the
combination of IA and clustering can have the constraining power as
strong as the the wide survey, as shown in table \ref{tab:constraints}
and figure \ref{fig:cosmo_PFS_Euclid}. If one combines all the three
probes in the deep survey, the constraints become stronger than the
conventional clustering analysis in the wide survey. We also show the
two-dimensional error contours of the cosmological parameters from the
wide survey in the bottom panel of figure \ref{fig:ogw0wacdm} which
can be compared to those from the deep survey in the top panel. These
results clearly demonstrate the importance of considering the IA and
kSZ effects.

\subsection{Choices of fiducial survey parameters}

The results of our Fisher matrix analyses in section \ref{sec:results}
and \ref{sec:deep_vs_wide} rely on the specific setup based on the
upcoming surveys.  Among several potential concerns in the actual
observations, the expected amplitude and error of kSZ and IA
statistics are less certain than those of galaxy clustering.
Specifically, the benefit of the IA statistics largely depends on the
fiducial setup of the parameters $\sigma_\gamma$ and $\AIA$, while
that of the kSZ statistics is affected by the choice of $\sigma_v$ and
$R_N$. In this subsection, we discuss the robustness of the benefit
combining the IA and kSZ data set with the galaxy clustering. To
elucidate this, allowing the parameters $\sigma_\gamma$, $\AIA$, and
$R_N$ (or $\sigma_v$) to vary, we estimate the FoM, defined by
equation (\ref{eq:FoM}).

Figure \ref{fig:FoM_PFS} shows the ratio of the FoM for the combined
data set of galaxy clustering and IA (or kSZ) to that for the galaxy
clustering alone, $\FoM_{g+E}/\FoM_{g}$ (or $\FoM_{g+v}/\FoM_{g}$).
The rightmost panels of the figure will be discussed in the next
subsection.  The upper panels plot the results for the geometric and
dynamical constraints, i.e., $\DA$, $H$ and $f\sigma_8$ at each
redshift slice. On the other hand, lower panels show the FoM for the
cosmological parameters.  As seen in the upper panels, the benefit of
combining kSZ and/or IA statistics increases with the number density
of galaxies, e.g., $10^4 n = 1.9, 6.0, 7.8, 3.1$, and $2.7$ $[h^3{\rm
    Mpc}^{-3}]$ at $z=0.7, 0.9, 1.3, 1.8$, and $2.2$ (see Table
\ref{tab:survey_pfs}).  Note that the results in the lower panels are
obtained by adding the CMB prior information, with the fluctuation
amplitude, $\sigma_{8,0}$, marginalized over.  Thus, the number of
cosmological parameters used to compute the FoM in equation
(\ref{eq:FoM}) is $N_p = 4,5,5$ and $6$ for the red, green, blue and
yellow curves, respectively.  As expected from the results in section
\ref{sec:constraints2}, the impact of combining IA or kSZ on the
improvement of cosmological parameters is more significant for the
models varying $\gamma$ and less significant for that varying
$\OmegaK$.  Even with the suppressed amplitude of ellipticity/velocity
fields or enhanced shape noise by a factor of 2, one can still expect
a fruitful benefit from the combination of galaxy clustering with
IA/kSZ. In particular, adopting the $w_0w_a\gamma$ non-flat model, the
improvement on each parameter reaches $\sim 20\%$, compared to the
case with galaxy clustering data alone.

\subsection{Effect of line-of-sight structures on kSZ statistics}\label{sec:los_ksz}

In this paper, as in previous works \cite[e.g.,][]{Hand:2012}, we
considered that the kSZ effect is observed in a three-dimensional
space, and statistical properties of the measured velocity fields are
described by the three-dimensional matter power spectrum through
equations (\ref{eq:Pij}) with (\ref{eq:kv}).  However, the
contribution of the kSZ effect to CMB anisotropies is in general given
by a line-of-sight integral of the velocity field.  Thus, unless we
use massive galaxy groups or clusters as a tracer of the velocity
field, the measured kSZ signals would be affected by other velocity
components arising predominantly from diffuse and extended sources
that may not fairly trace the large-scale matter flow, hence leading
to a suppression of the three-dimensional power spectra
\cite{Hernandez-Monteagudo:2015}.  To see this effect, we approximate
the impact of the line-of-sight integral by introducing a
multiplicative Gaussian smoothing kernel with the typical correlation
length $\slos$, $\Wksz(k_\parallel;\slos)=e^{-k_\parallel^2
  \slos^2/2}$.  The kSZ distortion field, $\delta T(\bfk)$, is then
modulated as $\delta T(\bfk) \rightarrow \delta
T(\bfk)\Wksz(k_\parallel;\slos)$. Accordingly, the power spectra that
include the velocity field, $\Pgv(\bfk), \PvE(\bfk)$ and $\Pvv(\bfk)$,
are modulated as $\Pgv(\bfk)\Wksz(k_\parallel;\slos)$,
$\PvE(\bfk)\Wksz(k_\parallel;\slos)$, and
$\Pvv(\bfk)\Wksz^2(k_\parallel;\slos)$, respectively.  It is not
trivial how the line-of-sight structure affects the velocity
dispersion, $\sigma_d^2$, which appears in the shot noise contribution
(see equation (\ref{eq:Pvv_with_shot_noise})).  Although such a
structure may introduce additional noise contribution to the one
modeled by equation (\ref{eq:sigma_d}), we assume for simplicity that
the velocity dispersion remains unchanged, as we have already seen the
impact of the increased velocity dispersion on FoM in left panels of
figure \ref{fig:FoM_PFS}.

The rightmost panels of figure \ref{fig:FoM_PFS} show the ratio of the
FoM for the combined data set of galaxy clustering and kSZ to that for
the galaxy clustering alone, $\FoM_{g+v} /\FoM_g$, as a function of
the smearing length, $\slos$. Note that in estimating the FoM, we
consider that the damping function $G$ is not a properly modeled
factor, and for a conservative estimate, we do not take into account
the AP effect of this function.  The fractional gain of the FoM by
adding kSZ decreases with increasing $\slos$, as expected. However,
even with such a conservative setting, we can still expect $5 - 10\%$
improvements at typical values of $\slos$, $\slos\sim 40-60 ~\himpc$.

As another example, let us also consider the case where the velocity
dispersion including the diffuse/extended components is modeled by the
line-of-sight integral just like the kSZ power spectra themselves.  In
such a case, the expression of the velocity dispersion in equation
(\ref{eq:sigma_d}) is modulated as
\begin{align}
\sigma_d^2 &= \int\frac{d^3q}{(2\pi)^3}\frac{\mu^2 P_{\theta\theta}(q;z)}{q^2} \Wksz^2(q\mu;\slos) \nn \\
&= \frac{1}{6\pi^2} \int^\infty_0 dq P_{\theta\theta}(q)
\left\{ -\frac{3e^{-q^2\slos^2}}{2q^2\slos^2} + \frac{3\sqrt{\pi}{\rm erf}(q\slos)}{4q^3\slos^3} \right\} \nn \\
&= \frac{1}{6\pi^2}\int^\infty_0 dq P_{\theta\theta}(q) 
\left( 
1 - \frac{3}{5} q^2\slos^2 + \frac{3}{14}q^4\slos^4 - \cdots 
\right),
\end{align}
where ${\rm erf}(x)$ is the error function and the third equality is
derived by the Taylor expansion. Adopting the estimation of the
velocity dispersion given above, we find that the fractional gain of
adding the kSZ effect is almost unchanged, a few per cent, at
$\slos=50\himpc$, compared to the undamped case ($\slos=0\himpc$).

Throughout this paper, we have considered the ``homogeneous'' kSZ
effect, which arises when the reionization process is complete
\cite{Sunyaev:1980,Ostriker:1986}.  However, on top of that, there is
a residual kSZ effect due to the ``patchy'' (or inhomogeneous)
reionization, which arises during the process of reionization, from
the proper motion of ionized bubbles around emitting sources, and it
can be an additional source of the noise for the kSZ signal
\cite{Aghanim:1996}. The contribution of the patchy kSZ effect becomes
significant at small scales, $\ell\sim k\chi > 1000$
\cite{Aghanim:1996,Planck-Collaboration:2016b}, while our analysis
focuses the data only up to quasi-nonlinear scales, $k\leq 0.2\hmpci$.
Thus in our analysis we safely ignore this effect.  However, when we
perform a more aggressive analysis including higher-$k$ modes, this
patchy reionization effect needs to be properly taken into account.

\subsection{Contribution of gravitational lensing to IA statistics}\label{sec:lensing}

So far, we have considered the observation of IA as one of the
cosmological probes ignoring the lensing effect. In principle, the
shape of the galaxies, projected onto the sky, can be very sensitive
to the lensing effect, and has been extensively used to detect and
measure the cosmic shear signals. This implies that unless properly
modeling it, the lensing effect on the E-mode ellipticity may be
regarded as a potential systematics that can degrade the geometric and
dynamical constraints. Nevertheless, one important point in the
present analysis using the IA is that, in contrast to the conventional
lensing analysis, one gets access to the cosmological information from
the three-dimensional power spectrum. In this subsection, we discuss a
quantitative impact of the lensing contribution on the observations of
IA, particularly focusing on the three-dimensional power spectrum of
E-mode ellipticity.

In the presence of the lensing effect, the observed E-mode ellipticity
defined in the three-dimensional Fourier space, $\gamma_{\rm
  E}(\bfk;z)$, is divided into two pieces, $\gamma_{\rm E}=\gamma_{\rm
  E}^{\rm (I)} + \gamma_{\rm E}^{\rm (GL)}$.  Here the former is
originated from the IA, and the latter represents the lens-induced
ellipticity. Then the (auto) E-mode power spectrum measured at a
redshift $z$ is expressed as
\begin{align}
 \PEE(\bfk;z)=\PEE^{\rm (I)}(\bfk;z)+\PEE^{\rm (GL)}(\bfk;z). 
\label{eq:pk_EE_IA_lens}
\end{align}
Note that in principle, there exists the cross talk between IA and
lensing, i.e., the gravitational shear--intrinsic ellipticity
correlation. However, such a cross talk becomes non-vanishing only if
we take the correlation between different $z$-slices. Since the
geometric and dynamical constraints considered in this paper are
obtained from individual $z$-slices, the relevant quantity to be
considered is only the E-mode lensing spectrum, $\PEE^{\rm (GL)}$.

\begin{figure*}[t]
\centering
\includegraphics[width=0.99\textwidth]{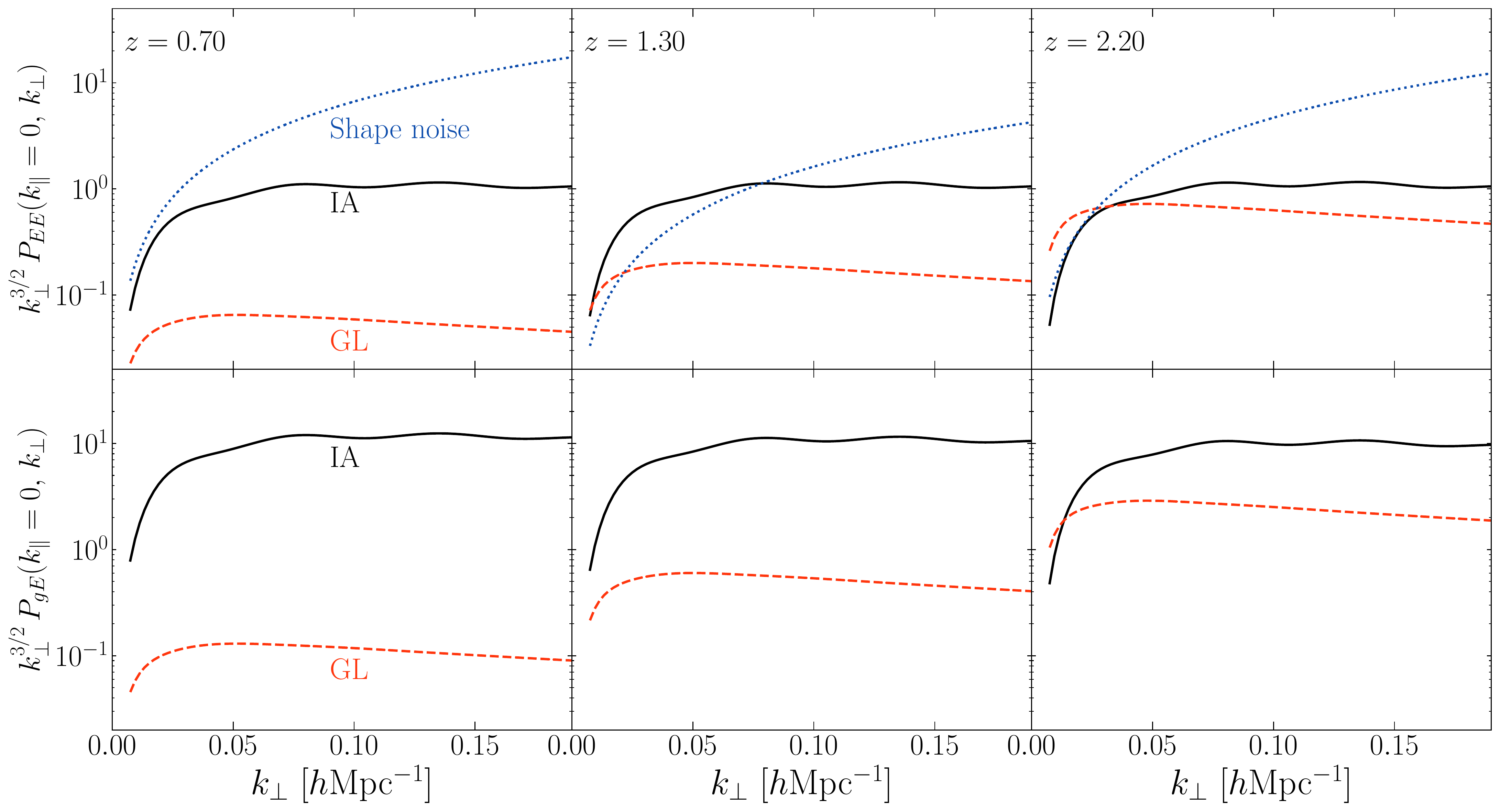}
\caption{Impact of lensing effects on the auto power spectra of E-mode
  ellipticity $\PEE$ (upper row) and cross power spectra between
  density and E-mode ellipticity $\PgE$ (lower row), at $z=0.7$ ({\it
  left}), $1.30$ ({\it middle}), and $2.20$ ({\it right}), which are
the lowest, central and highest redshift bins of the PFS survey,
respectively.  To highlight the most significant lensing impact, the
power spectra shown here are the results with $k_\parallel=0$, and the
results multiplied by $k_\perp^{3/2}$ are plotted as function of the
transverse wavenumber, $k_\perp$.  The red-dashed lines represent the
lensing contributions (i.e., $\PEE^{\rm (GL)}$ and $\PmE^{\rm(GL)}$ in
upper and lower panels, respectively), which are computed from
equations (\ref{eq:pk_EE_lens}) and (\ref{eq:pk_mE_lens}), adopting
the top-hat survey window ($\PEE^{\rm (GL)}$).  The redshift bin size,
$\Delta z = z_{\max}-z_{\min}$, is $\Delta z=0.2$ for $z=0.7$ and
$1.3$, and $\Delta z=0.4$ for $z=2.2$.  The black-solid lines are the
un-lensed power spectra, $\PEE^{\rm(I)}$ (upper) and $\PgE^{\rm(I)}$
(lower), originated purely from the IA and clustering. In upper
panels, we also show the non-vanishing noise contribution (see
equation (\ref{eq:PEE_with_shot_noise})), depicted as blue-dotted
lines.}
\label{fig:pk_EE_lens_IA}
\end{figure*}

Similarly, the observed density field is altered by gravitational
lensing, known as the magnification effect.  By denoting the observed
galaxy density field as $\delta_\obs$, one can decompose it into the
intrinsic density and the term due to magnification, $\delta_\obs =
\delta_g+\delta_\mu$.  Then the cross power spectrum between galaxy
density and ellipticity fields, $\PgE$, is expressed as
\begin{align}
 \PgE(\bfk;z)=\PgE^{\rm (I)}(\bfk;z)+\PmE^{\rm (GL)}(\bfk;z), 
\label{eq:pk_EE_IA_lens}
\end{align}
where the first term is the cross power spectrum between intrinsic
density and ellipticity fields considered so far, and the second term
represents the lens-induced cross-power spectrum.  Again, there are
also cross-talk terms, the galaxy density--lensing shear $\PgE^{\rm
  (GL)}$ and magnification--intrinsic ellipticity $\PmE^{\rm (I)}$
correlations.  Furthermore, the lens-induced ellipticities would be
correlated with the kSZ, leading to a non-zero contamination to
$P_{vE}$.  Since we consider the correlation functions in individual
$z$-slices, these cross-talks are negligible in our analysis.

Under the Limber approximation, $\PEE^{\rm (GL)}$ and $\PmE^{\rm
  (GL)}$ are analytically expressed as an integral of the comoving
distance \citep[e.g.][]{Matsubara:2000,Hui:2008}:
\begin{align}
\PEE^{\rm (GL)} (\bfk;\,z)  = 
& \left(\frac{3}{2}\frac{\Omegam H_0^2}{c^2}\right)^2 
|W_\parallel(k_\parallel)|^2  \nn \\
& \times
\int_0^\infty d\chi'\bigl\{w\bigl(\chi';\chi(z)\bigr)\bigr\}^2 
\Bigl\{\frac{\chi(z)}{\chi'}\Bigr\}^2 \nn \\ 
& \ \ \ \ \ \ \ \ 
\times \Plin\left(\frac{\chi(z)}{\chi'}k_\perp;\,z(\chi')\right),
\label{eq:pk_EE_lens} 
\\
\PmE^{\rm (GL)}(\bfk;z) = & 
2(\alpha_s -1) \PEE^{\rm (GL)}, \label{eq:pk_mE_lens}
\end{align}
where $\alpha_s$ is the logarithmic slope of the cumulative galaxy
luminosity function and the lensing kernel $w(\chi';\chi)$ is given by
$
w(\chi';\chi)=(\chi-\chi')\chi'/\{a(\chi')\,\chi\}\,\Theta(\chi-\chi')$
with $\Theta(x)$ being the Heaviside step function.  The function
$W_\parallel(k_\parallel)$ is the Fourier counterpart of the survey
window function along the line-of-sight direction,
$W_\parallel(x_\parallel)$. Equation (\ref{eq:pk_mE_lens}) coincides
with equation (15) of Ref. \cite{Hui:2008}, ignoring the transverse
survey window function $W_\perp$. Since our analysis targets
spectroscopic surveys with an accurate redshift determination provided
for each sample, we assume a top-hat window function,
\be
W_\parallel(x_\parallel) = 1/\sqrt{L}, \ \ \ \text{if} \ \  \bar{\chi}-L/2 < x_\parallel < \bar{\chi}+L/2,
\ee
and $W_\parallel(x_\parallel) = 0$ otherwise. Here $L$ is the radial
comoving size which corresponds to the redshift bin, given by
$L=\chi(z_{\max})-\chi(z_{\min})\simeq (z_{\max}-z_{\min})\, c/ H(z)$
(see table \ref{tab:survey_pfs} for the values of $z_{\max}$ and
$z_{\min}$ for each redshift bin). This top-hat window leads to
$|W_\parallel(k_\parallel)|^2 = (4/Lk_\parallel^2)\{\sin{(k_\parallel
  L / 2)}\}^2$ in Fourier space. This means that the lensing
contribution becomes maximum at $k_\parallel\ll1$, yielding
$|W_\parallel|^2\sim L$.

Figure \ref{fig:pk_EE_lens_IA} shows $\PEE$ (upper row) and $\PgE$
(lower row) at $z=0.7$, $1.30$, and $2.20$, which are the lowest,
central and highest redshift bins of the PFS survey, respectively.
The power spectra shown here are the results with $k_\parallel=0$ to
highlight the maximum lensing contributions, $\PEE^{\rm (GL)}$
(eq. \ref{eq:pk_EE_lens}) and $\PmE^{\rm (GL)}$
(eq. \ref{eq:pk_mE_lens}).  As increasing $z$, the amplitude of
$\PEE^{\rm (GL)}$ depicted as red dashed lines in the upper row, gets
larger.  However, apart from the shape noise, the signal coming from
the IA always dominates the E-mode power spectrum. Furthermore, the
amplitude of $\PEE^{\rm (GL)}$ is always smaller than the shape noise
expected from our fiducial setup of $\sigma_\gamma=0.2$, depicted as
blue dotted lines.  On the other hand, for the power spectrum $\PgE$,
the amplitude is controlled by the additional parameter $\alpha_s$
(eq. (\ref{eq:pk_mE_lens})), which depends on magnitude and redshift
of a given galaxy sample.  We adopt the typical values of $\alpha_s$,
$\alpha_s = 2,2.5,$ and $3$ for $z=0.7$, $1.30$, and $2.20$,
respectively \cite[e.g.,][]{Joachimi:2011}. Due to the extra redshift
dependence on $\alpha_s$, the lensing contribution to $\PmE^{\rm
  (GL)}$ increases faster toward higher $z$ that to $\PEE^{\rm
  (GL)}$. Nevertheless, the lensing contribution is still subdominant,
and we can clearly detect the BAO signal even for the case of
$k_\parallel=0$.

Taking the lens-induced $E$-mode ellipticity and galaxy density fields
to be systematic errors, we have repeated the Fisher matrix analysis,
for which the lens-induced auto power spectrum of $E$-mode and cross
power spectrum of magnification and $E$-mode are included in the
covariance at Eq.~(\ref{eq:covariance}). We then confirmed that the
changes in the estimated errors are negligibly small.  Furthermore,
instead of the top-hat filter function, we have examined another
filter, the Gaussian window function, given by
$|W_\parallel(k_\parallel)|^2=\sqrt{4\pi\Sigma^2}\,\exp(-k_\parallel^2
\Sigma^2)$ in Fourier space.  If we assume $\Sigma = L/\sqrt{4\pi}$,
the contribution becomes almost equivalent to the case with the
top-hat window \cite{Hui:2008}.  If we choose a wider window, e.g.,
$\Sigma = L$, the amplitude of the lens-induced power spectrum becomes
$\sqrt{4\pi}\sim 3.5$ times larger.  Even in that case, changes in the
statistical error on each parameter are still negligible, $< 1\%$,
namely at most the last digits of the values quoted in table
\ref{tab:constraints} are modulated.

Hence, we conclude that the lensing effect on the observations of the
IA gives a sub-dominant contribution to $\PEE$ and $\PgE$ as long as
we consider spectroscopic surveys, and it hardly changes the
cosmological constraints.


\section{Conclusions \label{sec:conclusion}}

In this paper, based on the Fisher matrix analysis, we have shown that
combining IA and kSZ statistics with the conventional galaxy
clustering statistics substantially improves the geometric and
dynamical constraints on cosmology.  As a representative of deep
galaxy surveys for the forecast study, we considered the Subaru PFS,
whose angular area perfectly overlaps with those from the HSC survey
and the CMB-S4 experiment.  We found that even without the galaxy
clustering, observations of IA and kSZ enable us to constrain $\DA$
and $H$, with the achievable precision down to a few percent. This
demonstrated that constraining the geometric distances with kSZ and IA
effects would help addressing recent systematics-related issues such
as the Hubble tension.

For cosmological parameter estimations, a relative merit of adding kSZ
and IA statistics to the galaxy clustering depends on cosmological
models. We found that the improvement of combining kSZ and IA to
clustering statistics is maximized if we simultaneously constrain the
time-varying dark energy equation-of-state parameter
$w(a)=w_0+(1-a)w_a$ and the growth index $\gamma$ characterizing the
modification of gravity in a non-flat universe ($w_0w_a\gamma$
non-flat model). In such a model, with the CMB prior information from
the Planck experiment, the PFS-like deep survey is shown to improve
the constraints by $31\%$ for $\gamma$ and $> 35\%$ for others except
the prior-dominated constraint on $\OmegaK$.

To see the gain of adding IA and kSZ for a different survey setup, we
have also performed the Fisher matrix analysis for the {\it
  Euclid}-like wide galaxy survey, whose survey area is partly
overlapped by half with the CMB-S4 experiment on the sky.  Due to the
large volume, such a wide survey can give tighter constraints on $f$,
$\DA$ and $H$ at each redshift bin.  However, when considering the
cosmological models which vary the growth index parameter, a deep
survey is more effective than a wide survey, and can get tighter
constraints. As a result, in the $w_0w_a\gamma$ non-flat model, by
combining kSZ and IA measurements with the clustering measurement,
cosmological constraints from the PFS-like deep survey can be tighter
than those with the conventional clustering-only measurement from the
{\it Euclid}-like wide survey.  Finally, we have also discussed the
potential impact of the lensing effect on the observation of IA and
line-of-sight structures on the kSZ statistics, the former of which
can systematically change the IA auto-power spectrum, $\PEE$ (see
equation (\ref{eq:pk_EE_IA_lens})). However, even for the deep survey
considered, the lens-induced ellipticity is shown to give a negligible
contribution as long as we consider the three-dimensional power
spectrum, and hence the cosmological parameter estimated from the IA
data is hardly changed. For the kSZ statistics, even with a large
correlation length of $D_{\rm LOS}\sim40-60\,h^{-1}$\,Mpc, the impact
of the line-of-sight structures on the cosmological parameters is
fairly small as long as we consider a joint analysis with the galaxy
clustering.

\begin{table*}[bt!]
\caption{Same as table \ref{tab:constraints} but one-dimensional
  fractional marginalized errors on cosmological parameters,
  $\sigma/\theta^\fid$, when only the data up to the linear scales
  $k_{\max}=0.1\,\hmpci$ are used.  }
\begin{center}
\begin{tabular}{l c c  c c c c c c c c c  }
\hline \hline
& & \ \ \ & \multicolumn{4}{c}{Deep survey} & \ \ \ \ & \multicolumn{4}{c}{Wide survey}  \\
&&& Clustering &  &  & &  & Clustering &  &  & \\
Model & $\sigma/\theta^\fid$ &&    only       & +kSZ         & +IA            & +IA+kSZ &&    only   & +kSZ        & +IA            & +IA+kSZ \\
\hline
$w_0,w_a$ & $\Omegam$ && 0.163 & 0.142 & 0.138 & 0.128 & & 0.117 & 0.111 & 0.106 & 0.105 \\
flat & $w_0$ && 0.452 & 0.415 & 0.404 &  0.377  && 0.327 & 0.317 & 0.309 & 0.303  \\
& $w_a$ && 1.162 & 1.047 & 1.021 & 0.952 && 0.857 & 0.833 & 0.813 & 0.795  \\
& $H_0$ && 0.0785 & 0.0679 & 0.0657 & 0.0608 && 0.0538 &	 0.0509 & 0.0484 & 0.0475  \\
\hline
$w_0,w_a$ &$\Omegam$ &&0.185 & 0.163 & 0.158 & 0.145&& 0.128 & 0.124 & 0.117 & 0.115  \\
 non-flat & $w_0$ && 0.518 & 0.465 & 0.452 & 0.415 && 0.371 & 0.358 & 0.344 & 0.337  \\
& $w_a$ && 1.377 & 1.206 & 1.175 & 1.077 && 0.964 & 0.928 & 0.896 & 0.874  \\
& $H_0$ && 0.0912 & 0.0805 & 0.0777 & 0.0715 && 0.0626 & 0.0607 & 0.0568 & 0.0560  \\
& $100\OmegaK$ && 0.394 & 0.382 & 0.375 & 0.368 && 0.233 & 0.233 & 0.227 & 0.226  \\
\hline
$w_0,w_a,\gamma$ &$\Omegam$ && 0.277 & 0.187 & 0.179 & 0.150 && 0.170 & 0.152 & 0.135 & 0.126  \\
flat& $w_0$ && 0.786 & 0.557 & 0.535 & 0.447 && 0.485 & 0.446 & 0.400 & 0.368  \\
& $w_a$ && 1.862 & 1.359 & 1.310 & 1.107 && 1.222 & 1.140 & 1.032 & 0.948  \\
& $H_0$ && 0.1360 & 0.0910 & 0.0867 & 0.0721 && 0.0795 & 0.0708 & 0.0623 & 0.0580  \\
& $\gamma$ && 0.499 & 0.375 & 0.370 & 0.326 && 0.349 & 0.323 & 0.300 & 0.259  \\
\hline
$w_0,w_a,\gamma$ & $\Omegam$ && 0.345 & 0.217 & 0.207 & 0.169 && 0.243 & 0.208 & 0.168 & 0.151  \\
non-flat& $w_0$ && 0.986 &  0.633& 0.605  & 0.492 && 0.718 & 0.616 & 0.507 & 0.447  \\
& $w_a$ && 2.410 & 1.583 & 1.521 & 1.248 && 1.792 & 1.547 & 1.291 & 1.133 \\
& $H_0$ && 0.1722 & 0.1080 & 0.1027 & 0.0838 && 0.1212 & 0.1033 & 0.0834 & 0.0747 \\
& $100\OmegaK$ && 0.439 & 0.392 & 0.384 & 0.371 && 0.305 & 0.285 & 0.259 & 0.247 \\
& $\gamma$ && 0.559 & 0.386 & 0.380 & 0.329 && 0.459 & 0.397 & 0.343 & 0.284  \\
\hline\hline
\end{tabular}
\label{tab:constraints_kmin010}
\end{center}

\end{table*}

In this paper, focusing specifically on the measurements of geometric
and dynamical distortions, we have shown that the combination of both
IA and kSZ with galaxy clustering is beneficial. Note, however, that
the present analysis using only the BAO and RSD information is not as
powerful as that using the full shape of the the underlying matter
power spectrum.  Although one advantage in the present analysis is
that the systematics arising from the nonlinearity is less severe and
hence conservative, it would be highly desirable for more tighter
cosmological constraints, in particular on the neutrino masses, to
make use of the full shape information \cite{Boyle:2018}. Indeed, the
analysis with the full shape of the power spectrum has been performed
in the conventional galaxy clustering analysis
\cite{Okumura:2008,Ivanov:2020,dAmico:2020,Philcox:2020,Kobayashi:2022}.
However, the analysis with full-shape information needs a proper
nonlinear modeling, and compared to the modeling of the nonlinearities
for clustering statistics, less studies have been made for velocity
and ellipticity statistics
\cite[e.g.,][]{Okumura:2014,Blazek:2019,Vlah:2020}. Thus, before we
extend our joint analysis of density, velocity and ellipticity fields
to include the full-shape spectra, we need to develop models of
nonlinear power spectra for the velocity and ellipticity fields and
test them with numerical simulations.  Such analytical and numerical
analyses will be performed in future work.

\begin{acknowledgments}

We would like to thank Masahiro Takada, Ryu Makiya, Eiichiro Komatsu,
Naonori Sugiyama and Sunao Sugiyama for helpful discussions.
T.~O. also thanks the Subaru PFS Cosmology Working Group for useful
correspondences during the regular telecon. T.~O. acknowledges support
from the Ministry of Science and Technology of Taiwan under Grants
Nos. MOST 110-2112-M-001-045- and 111-2112-M-001-061- and the Career Development Award,
Academia Sinina (AS-CDA-108-M02) for the period of 2019 to 2023.
A.~T. was supported by MEXT/JSPS KAKENHI Grant Number JP17H06359,
JP20H05861 and JP21H01081. A.~T. also acknowledges financial support
from Japan Science and Technology Agency (JST) AIP Acceleration
Research Grant Number JP20317829.
\end{acknowledgments}

\appendix

\section{CMB prior}\label{sec:prior}

In this Appendix, we describe the CMB prior information added to the
Fisher matrix of the LSS probes (see figure \ref{fig:flowchart}). In
the analysis presented in this paper, the CMB prior information is
used to estimate the forecast constraints on cosmological parameters,
except for the minimal cosmological model ($w_0$ flat model).

First of all, our primary interest is how the geometric and dynamical
constraints derived from the BAO and RSD measurements can be used to
test cosmological models, with the power spectra of each LSS probe
characterized by Eq.~(\ref{eq:Pij_AP2}). For this purpose, we
specifically use the information determined mainly from the CMB
acoustic scales. We follow Ref.~\cite{Aubourg:2015} and use the
information on $\omegacb \equiv \Omegacb h^2$ and $\DM(1090)/r_{\rm
  d}$, fixing the energy density of neutrinos $\omega_\nu$ and baryon
$\omegab$ respectively to $\omega_\nu = 6.42\times 10^{-4}$ and
$\omegab=0.022284$, the former of which corresponds to the total mass
of $\sum{m_\nu} = 0.06$ [eV]. Here, the $\Omegacb$ is the density
parameter of CDM and baryons, i.e., $\Omegacb = \Omegac +
\Omegab$. The quantity $\DM(z) = (1+z)\DA(z)$ is the comoving angular
diameter distance \cite{Hogg:1999}, and $r_d$ is the sound horizon at
the drag epoch, for which we use the numerically calibrated
approximation:
\be
r_{\rm d} \simeq \frac{55.124\exp{\left[ -72.3(\omega_\nu+0.0006)^2 \right]}}{\omegacb^{0.25351}\omegab^{0.12807}} ~{\rm Mpc},
\ee
with $\omega_\nu$ and $\omegab$ kept fixed to the values mentioned
above.  Ref.~\cite{Aubourg:2015} found that the acoustic scale
information on the data vector
$\Theta_\alpha=(\omegacb,\DM(1090)/r_{\rm d})$ can be described by a
Gaussian likelihood with mean and covariance (see also
\cite{Ivanov:2020})\footnote{To be precise, Ref.~\cite{{Aubourg:2015}}
  provided a Gaussian likelihood for the three parameters
  $\Theta_\alpha=(\omegab,\, \omegacb,\,\DM(1090)/r_{\rm d}$) having
  the $3\times3$ covariance matrix. Since we consider $\omegab$ to be
  fixed, the relevant prior information is described by the $2\times2$
  covariance matrix given at equation (\ref{eq:CMB_prior_mean_cov}).}:
\begin{align}
& \mu_{\Theta} = \left( 0.1386, \, 94.33 \right), \\
& C_{\Theta} = \begin{pmatrix} 7.452\times 10^{-6} & -3.605 \times 10^{-5}  \\ -3.605 \times 10^{-5} & 0.004264 \end{pmatrix} .
 \label{eq:CMB_prior_mean_cov}
\end{align}
The inverse of this error matrix is the Fisher matrix, $\bfF_{{\rm
    CMB}} = \bfC_{\Theta}^{-1}$, shown in the lower left of the
flowchart in figure \ref{fig:flowchart}.  It is then converted to the
Fisher matrix for a given cosmological model of interest, $\bfS_{{\rm
    CMB}}$, through equation (\ref{eq:projection_fisher}). We have
also tried another CMB prior used in our early study
\cite{Taruya:2020}, based on Seo \& Eisenstein \cite{Seo:2003}, and
confirmed that our forecast results almost remain unchanged.

\section{Forecast results with the conservative cutoff of $k_{\max}=0.1h{\rm Mpc}^{-1}$}
\label{sec:conservative}


\begin{figure}[tb]
\centering
\includegraphics[width=0.496\textwidth]{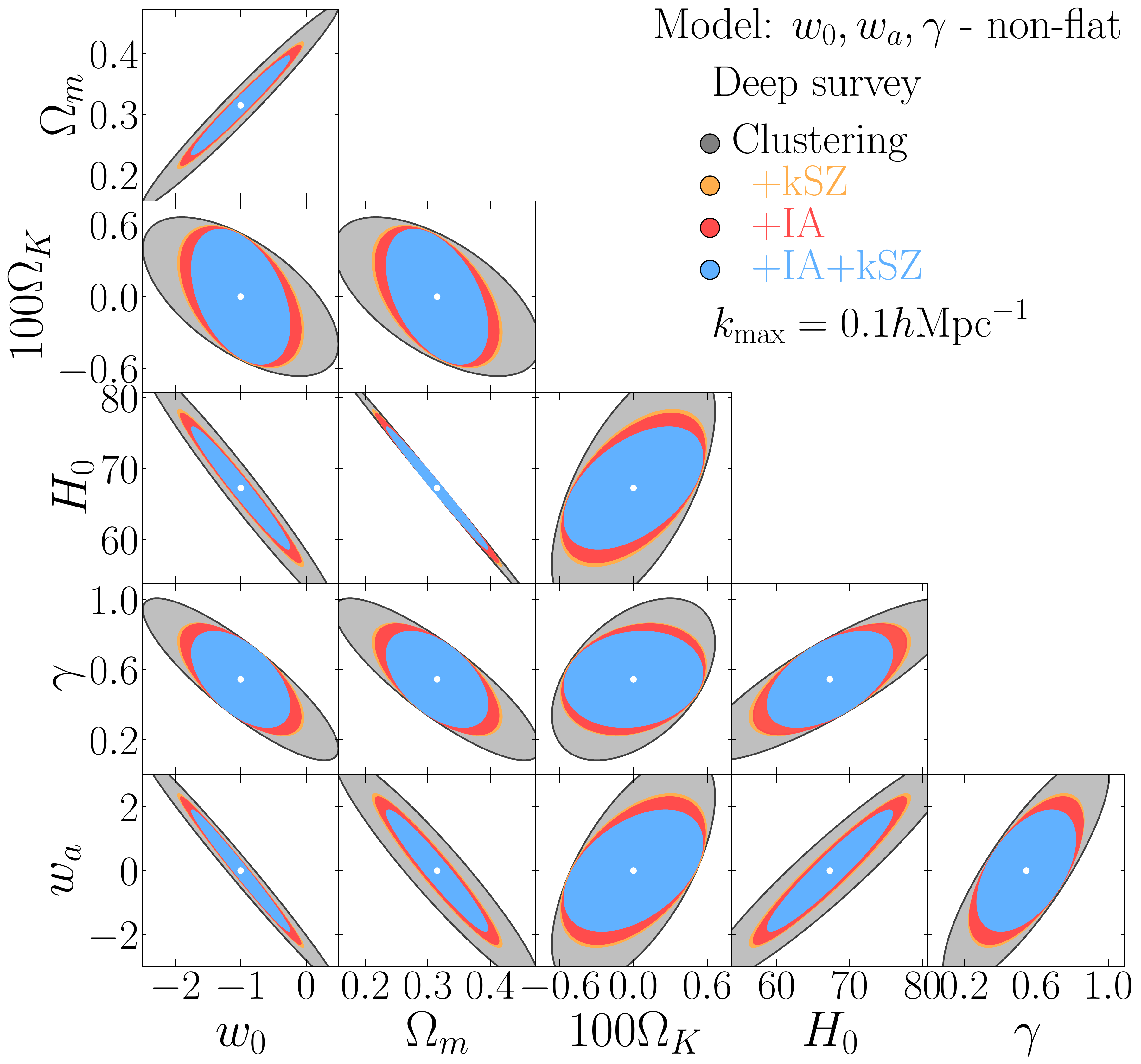}
\includegraphics[width=0.496\textwidth]{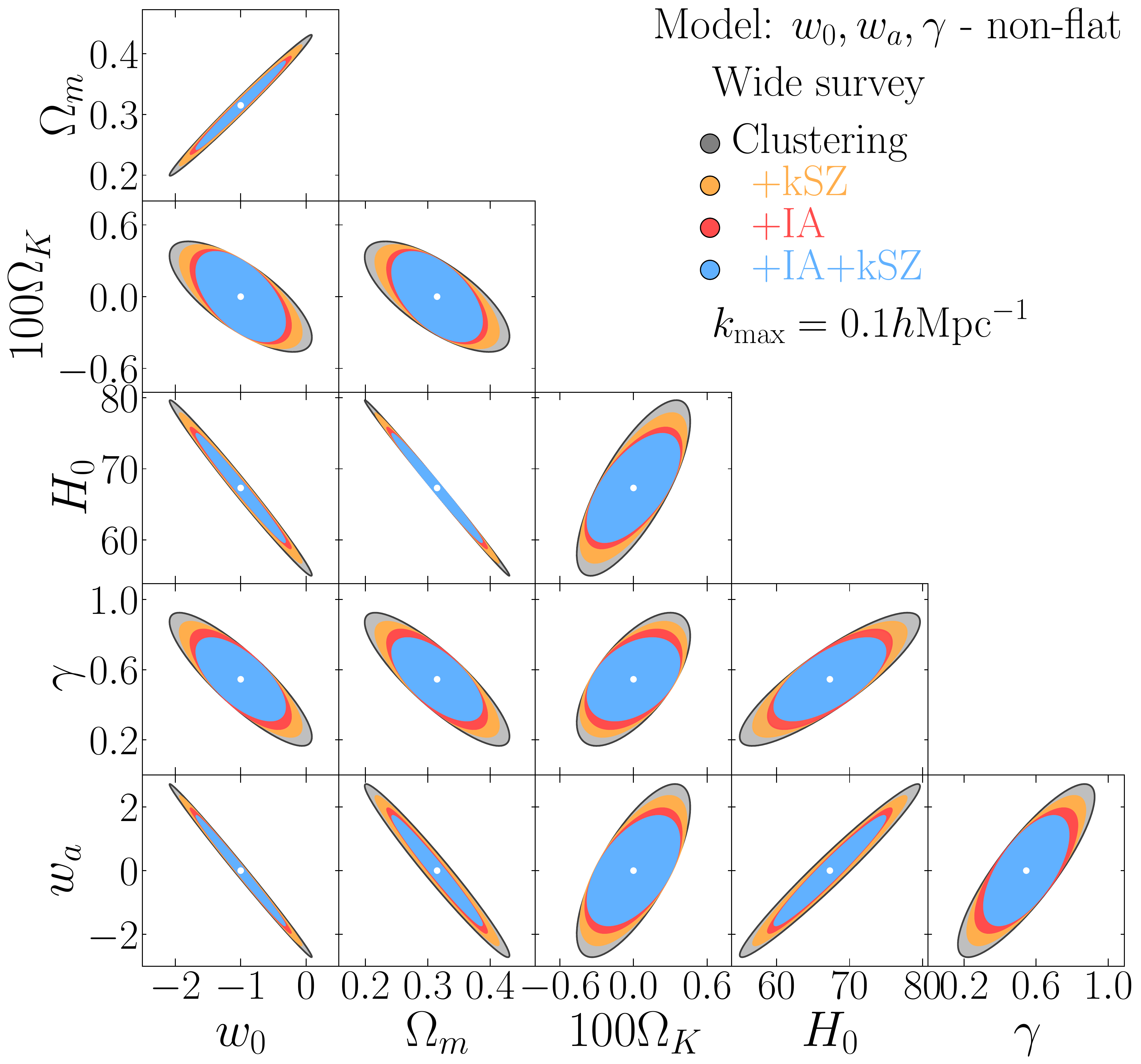}
\caption{ Cosmological constraints on the $w_0w_a\gamma$ non-flat
  model from the deep ({\it Top }) and wide ({\it Bottom}) surveys.
  These results are similar with figure \ref{fig:ogw0wacdm} but using
  the conservative range of data of $k_{\max}=0.1\,\hmpci$.  }
\label{fig:ogw0wacdm_kmin010}
\end{figure}

In sections \ref{sec:results} and \ref{sec:discussion}, forecast
constraints on cosmological parameters, including the geometric
distances and growth of structure, were derived focusing on the
upcoming deep and wide galaxy surveys, PFS and {\it Euclid},
respectively. In doing so, one important assumption was that the
linear theory template for the power spectra is applicable to the
weakly nonlinear scales, setting the maximum wavenumber to $k_{\rm
  max}=0.2\,h$\,Mpc$^{-1}$ for all the three LSS probes. While our
analysis is still conservative in the sense that we only use the
geometric and dynamical information obtained from the BAO and RSD
measurements, restricting the data to the linear scales of $k \leq
0.1\, h\,{\rm Mpc}^{-1}$ would yield a more conservative and robust
forecast results, and no intricate modeling of the nonlinear
systematics needs to be developed. In this appendix, repeating the
Fisher matrix analysis but with $k_{\rm max}=0.1\,h$\,Mpc$^{-1}$, we
summarize the forecast constraints on cosmological parameters.

First we consider the deep survey.  The left half of table
\ref{tab:constraints_kmin010} summarizes the one-dimensional
marginalized errors on cosmological parameters, which are compared to
results with $k_{\rm max}=0.2\,h$\,Mpc$^{-1}$ listed in the left half
of table \ref{tab:constraints}. The results are also shown visually as
the hollow bars in figure \ref{fig:cosmo_PFS_Euclid}.  The expected
errors obtained from the clustering-only analysis with $k_{\rm
  max}=0.1\,h$\,Mpc$^{-1}$ are roughly twice as large as those with
$k_{\rm max}=0.2\,h$\,Mpc$^{-1}$.  Interestingly, however, the
fractional gain of the cosmological power by adding the kSZ and/or IA
measurements is more significant for the conservative analysis with
$k_{\max}=0.1\,h\,{\rm Mpc}^{-1}$.  For instance, in the most general
model considered in this paper, namely the $w_0w_a\gamma$ non-flat
model (see table \ref{tab:models}), the improvements by $48\%$ and
$41\%$, relative to the clustering-only analysis are respectively
achieved for the constraints on $w_a$ and $\gamma$. These are compared
to the relative improvements by $35\%$ and $31\%$ in the cases with
$k_{\max}=0.2\,h\,{\rm Mpc}^{-1}$.

Let us then compare the forecast results for the deep survey with
those for the wide galaxy survey. As seen in the right side of table
\ref{tab:constraints_kmin010} (see also the hollow bars in figure
\ref{fig:cosmo_PFS_Euclid}), the constraining power of the
clustering-only analysis from the wide survey is $25-40\%$ stronger
than that from the deep survey. This is more or less the same as the
case with the aggressive cutoff of $k_{\max}=0.2\,\hmpci$.  However,
one notable point is that the benefit of combining the IA and kSZ
measurements is more significant for the deep survey than that for the
wide survey. In particular, in the $w_0w_a\gamma$ non-flat model,
combining either IA or kSZ with clustering in the deep survey can beat
the constraining power of the wide survey.  For illustration, in
figure \ref{fig:ogw0wacdm_kmin010}, the expected two-dimensional error
contours on the cosmological parameters are shown in the
$w_0w_a\gamma$ non-flat model.  This figure is similar with figure
\ref{fig:ogw0wacdm}, but here we adopt the conservative cut,
$k_{\max}=0.1\, h\, {\rm Mpc}^{-1}$, instead of $0.2\, h\, {\rm
  Mpc}^{-1}$. Clearly, the relative impact of combining IA and kSZ is
rather large for the deep survey, manifesting tighter constraints not
only on the growth index but also on other parameters including the
curvature parameter.

\bibliography{ms.bbl}
\end{document}